\documentclass[referee]{solarphysics}
\usepackage[optionalrh,solaromanenum]{spr-sola-addons} 
\usepackage[pdftex]{graphicx}                    
\usepackage{color}                       
\usepackage{url}                         
\usepackage{rotating}
\usepackage{sidecap}
\usepackage{setspace}
\usepackage{epstopdf}

\def\apj{Astrophys. J.} 
\def\apjl{Astrophys. J. Lett.} 
\def\apjs{Astrophys. J. Suppl.} 
\def\aap{Astron. Astrophys.}%
\def\aaps{Astron. Astrophys. Suppl.}%
\def\solphys{Solar~Phys.}%
\def\ssr{Space~Sci.~Rev.}%
\def\grl{Geophys.~Res.~Lett.}%
\def\sun{\hbox{$\odot$}}
\def\degr{\hbox{$^\circ$}}

\def\arcsec{\hbox{$^{\prime\prime}$}}

\begin{document}

\begin{article}

\begin{opening}

\title{Statistical Analysis of Large-scale EUV Waves Observed
by STEREO/EUVI}

%
\author{N.~\surname{Muhr}$^{1,2}$\sep
        A.~M.\surname{Veronig}$^{1}$\sep
        I.~W.\surname{Kienreich}$^{1}$\sep
        B.~\surname{Vr\v snak}$^{2}$\sep
        M.~\surname{Temmer}$^{1}$\sep
        B.~M.\surname{Bein}$^{1}$}

%
\runningtitle{Statistical Analysis of Large-scale EUV Waves}

%
  \institute{$^{1}$ IGAM/Kanzelh\"ohe Observatory, Institute of Physics, University of Graz,
              Universit\"atsplatz 5, A--8010 Graz, Austria\\
                     email: \url{muhrn@edu.uni-graz.at}\\
             $^{2}$ Hvar Observatory, Faculty of Geodesy, University of Zagreb,
              Ka\v{c}i\'{c}eva 26, HR--10000 Zagreb, Croatia
             }

\begin{abstract}
We present a statistical analysis of 60 strong large-scale EUV wave events that occurred during
January 2007 to February 2011 with the STEREO twin spacecraft
regarding their kinematical evolution and wave pulse
characteristics. For the start velocity, we obtain for the arithmetic mean $312\pm115$~km~s$^{-1}$ (within a range of 100$-$630~km~s$^{-1}$). For the mean (linear) velocity, the arithmetic mean is $254\pm76$~km~s$^{-1}$ (within a range of 130$-$470~km~s$^{-1}$). 52\% of all waves under study show a distinct deceleration during their propagation ($a\leq-50$~m~s$^{-2}$), the other 48\% are consistent with a constant speed within the uncertainties ($-50\leq a\leq50$~m~s$^{-2}$). The start velocity and the acceleration show a strong anticorrelation with $c\approx-0.8$, \textit{i.e.} initially faster events undergo stronger deceleration than slower events. The (smooth) transition between constant propagation for slow events and deceleration in faster events occurs at an EUV wave start velocity of $v\approx230$~km~s$^{-1}$, which corresponds well to the fast-mode speed in the quiet corona. These findings provide strong evidence that the EUV waves under study are indeed large-amplitude fast-mode MHD waves. This interpretation is further supported by the correlations obtained between the peak velocity and the peak amplitude, impulsiveness, and build-up time of the disturbance. We obtained the following association rates of EUV wave events to other solar phenomena: 95\% are associated with a coronal mass ejection (CME), 74\% to a solar flare, 15\% to interplanetary type II bursts, and 22\% to coronal type II bursts. These findings are consistent with the interpretation that the associated CMEs are the EUV waves' driving agent.
\end{abstract}

%

\end{opening}

\section{Introduction}
Globally propagating disturbances in the solar corona have been
first reported by \inlinecite{moses97} and \inlinecite{thompson98} in images
recorded by the Extreme Ultraviolet Imaging Telescope
(EIT; \opencite{delaboudiniere95}) aboard the \textit{Solar and Heliospheric
Observatory} (SoHO; \opencite{domingo95}) spacecraft, thereafter called
`EIT waves', or more generally, `EUV waves' or `large-scale coronal waves'. So far a hand full of statistical studies exist on large-scale coronal waves. \inlinecite{klassen00} analyzed 19 EUV waves reporting propagation speeds in the range of 170 -- 350~km~s$^{-1}$.
Comparable results were obtained by \inlinecite{thompson09} who studied 176 events that occurred in the solar minimum period from March 1997 to June 1998 with typical
velocities of 200--400~km~s$^{-1}$. They also tried to solve the question of the physical origin of EUV waves by examining the association rates to CMEs, flares and type II bursts. An unambiguous association of EUV waves with CMEs was found, significantly stronger than the association with flares and type II bursts (see also \opencite{biesecker02}; \opencite{cliver05}).

The detection of fast EUV waves may have been hampered by the time cadence of the EIT instrument ($\gtrsim 12$~min). These limitations were overcome by the launch of the \textit{STEREO (Solar-Terrestrial Relations
Observatory}; \opencite{kaiser08}) twin spacecraft in 2006 with its \textit{Extreme
UltraViolet Imager} (EUVI; \opencite{howard08}). The identically built instruments
(EUVI-A and EUVI-B) observe the solar corona in four different EUV
passbands with a cadence as high as 75~s and a large
field of view (FoV; up to 1.7~\textit{R}$_{\hbox{$\odot$}}$) from two different vantage
points. The first statistical study using EUVI observations was presented by \inlinecite{warmuth11}. These authors studied the kinematics of a set of 17 STEREO/EUVI and 61 SOHO/EIT wave observations, concluding that there are three different classes of EUV waves. Class 1 events are characterized by speeds higher than $320$~km~s$^{-1}$ and show deceleration during propagation. The deceleration is stronger in fast than in slow EUV waves, consistent with the nonlinear evolution of
large-amplitude waves or shocks. Class 2 events propagate with almost constant speeds in the range 170 -- 320~km~s$^{-1}$ (which is comparable to the local fast-mode speed in the solar corona) and can be interpreted as linear small-amplitude waves. The events of class 3 have velocities which are smaller than the coronal sound speed, and can thus not be integrated into the fast-mode wave interpretation. They may be either a result of magnetic reconfiguration \citep{delanee00, attrill07} or slow-mode waves \citep{podladchikova10}.

Further improvements on the observations of EUV waves came with the launch of the \textit{Solar Dynamics Observatory} (SDO; \opencite{pesnell12}) satellite in 2010. One of the instruments onboard SDO is the \textit{Atmospheric Imaging Assembly} (AIA; \opencite{lemen12}) that provides high-cadence (up to 12~s) EUV imagery with high sensitivity over a broad temperature range. \inlinecite{nitta13b} conducted a first AIA-based statistical study which consisted of 138 EUV wave events that occurred during April 2010 to January 2013. These authors studied the early propagation phase of the EUV waves in different propagation directions. For each event, the direction of the highest velocity was determined, giving a sample mean of $v\approx630$~km~s$^{-1}$, considerably higher than former wave studies using EIT or EUVI observations. In contrast to the \inlinecite{warmuth11} study, \inlinecite{nitta13b} found neither a correlation between the wave velocity and acceleration nor an evidence for a multi-class population of large-scale waves.

Two different interpretations for EUV waves are discussed, wave and non-wave models.
The wave model interprets the EUV transient as fast-mode MHD wave initiated either by the associated CME, flare or small scale ejecta \citep{vrsnak02,warmuth04a,veronig10,patsourakos12}. The wave model is supported by a variety of different observational findings: reflection, refraction and transmission of EUV waves across the boundaries of coronal holes or active regions \citep{thompson99,veronig08,gopalswamy09,shen12,olmedo12,li12,kienreich13,yang13}, the pulse broadening and amplitude decrease \citep{wills06,warmuth10,veronig10,long11,muhr11}, the downward plasma motions observed at the wave front \citep{veronig11,harra11}, the occurrence of homologous EUV waves showing a correlation between the wave speed and its amplitude \citep{kienreich11,zheng12} as well as the observations of multiple pulses \citep{liu12,liu14}.
Recent studies based on observations from SDO/AIA by \inlinecite{li12}, \inlinecite{cheng12} and \inlinecite{liu12} address also the wave nature of the EUV transient. In addition to the typical wave-like properties they observe the formation of secondary waves at active region boundaries and/or the separation of the diffuse wave signature from the driving CME at the end of the CME acceleration phase.

In the non-wave model, EUV wave signatures are interpreted as coronal ground track of successively opening and restructuring magnetic field lines caused by the expanding CME \citep{delanee99,chen02,attrill07,delannee08,dai10}. The observations of stationary wave fronts and the rotation of wave fronts \citep{delanee99,attrill07} are in support of the non-wave model.

This debate about the nature of the events resulted in a combination of the wave and the non-wave model, first suggested by \inlinecite{zhukov04}. MHD simulations by \inlinecite{cohen09} and \inlinecite{downs11} indeed revealed both wave as well as non-wave characteristics. The high cadence AIA imagery provided for the first time observations of simultaneously launched fast and slow EUV transients and support the combined wave/non-wave model \citep{liu10,chen11,cheng12,dai12}. For further discussions we refer to recent reviews by \inlinecite{gallagher10}, \inlinecite{wills10}, \inlinecite{warmuth10},
\inlinecite{patsourakos12} and \inlinecite{liu14}.

In this paper, we present a statistical analysis of EUV waves observed by STEREO/EUVI. The 60 EUV wave events under study occurred during
January 2007 to February 2011 which corresponds to the transition between the declining phase of solar cycle 23 and the rising phase of solar cycle 24, including the extreme solar minimum. This time range has two advantages: a) on average there is a small number of active regions which disturb the propagation of the EUV events under study; b) the downlink rate of the two STEREO spacecraft was still high enough to provide data at high cadence. The disadvantage, however, is that there is only a small number of strong EUV wave events included in the study, which occur preferentially during times of high solar activity. We study the kinematics
(velocity and acceleration) and wave pulse
characteristics (amplitude, impulsiveness) of the events as well as their association rates to CMEs, flares and type II bursts. Seven of the 60 events
were observed by both STEREO spacecraft and can thus be compared to study the differences in the obtained kinematics due to the observations from different vantage points.

A description of the data is given in Section~\ref{sec:datadescription} and the analysis technique
in Section~\ref{sec:analysis}. The
statistical results on the wave kinematics and the wave pulse
characteristics conducted by the semi-automated perturbation
profile method are presented in Section~\ref{sec:results}. A
discussion and conclusion on the results is given in
Section~\ref{sec:discussion}. In the Appendix of this paper we compare the results of kinematical analysis of a representative
sub-sample based on the visual tracking method \textit{vs.} perturbation profiles.

\section{Data and Observations}\label{sec:datadescription}
The EUVI instruments are part of the \textit{Sun Earth
Connection Coronal and Heliospheric Investigation}
(SECCHI; \opencite{howard08}) instrument suite on board the STEREO-A
(Ahead; ST-A) and STEREO-B (Behind; ST-B) spacecraft (s/c). The steady separation of
ST-A and ST-B is 45$\degr$ per year. EUVI observes the chromosphere and
low corona in four spectral channels (304$\hbox{\AA}$: dominant emission from He~{\sc ii} ion, T$\approx$
0.07 MK; 171$\hbox{\AA}$: Fe~{\sc ix}, T$\approx$ 1~MK;
195$\hbox{\AA}$: Fe~{\sc xii}, T$\approx$ 1.5~MK; 284$\hbox{\AA}$: Fe~{\sc xv}, T$\approx$ 2.25~MK) out to 1.7 \textit{R}$_{\hbox{$\odot$}}$ with
a pixel-limited spatial resolution of 1$\arcsec$.6 pixel$^{-1}$ \citep{wuelser04}.
EUV waves are best observed in the 195$\hbox{\AA}$ channel. In the data set under study, the EUVI 195~$\hbox{\AA}$ cadence was 2.5~min in 13 events, 5~min in 35 events, and 10~min in 12 events.

The statistical analysis started with a manual inspection of the EUVI data
set in the 195~$\hbox{\AA}$ passband to select suitable events.
To this aim, an online EUV event catalogue was established$\footnote{swe.uni-graz.at/stereo\_waves}$. In
this catalogue the identified EUV waves were classified and collected in event lists
for both EUVI instruments. For each event the date, start time, center position in the EUVI-A and EUVI-B field of view (FoV), and a (subjective) three-scale classification (1: weak, 2: moderate, 3: strong) are given.
Events of class 1 are weak in intensity and can only be observed for distances up to $\approx$80~Mm from the source region (comparable to the `mini CMEs' reported in \inlinecite{innes09}). Class 2 events are moderate in their intensity enhancement and can be observed over a maximum distance range up to $\approx$400 Mm. Class 3 events are strong in intensity and propagate over the whole solar disk.

During the time period from January 2007 to
February 2011 more than 1000 EUV transients were identified in the STEREO/EUVI data. The number of events belonging to the three different classes is as follows: class 1 $\approx$ 800 events, class 2 $\approx$ 150 events, and class 3 $\approx$ 50 events which means that the overall occurrence rate for class 1 events is 80\%, for class 2 events it is 15\% and for class 3 events it is 5\%.

We excluded all events with importance class 1 and faint class 2 events as well as
those events having the wave center located behind the
visible solar disk or very close to the solar limb (corresponding to events located within 75$\degr$ up to 90$\degr$ from the central meridian) in order to avoid large errors in the wave kinematics due to projection effects. In summary, a total of 60 different EUV wave events (most of them is of class 3 and some events are from class 2) is analyzed (36 in
ST-A and 31 in ST-B). 8 events were observed and analyzed in both STEREO s/c.

For the events under study we determined the CME, flare and type II burst association rate by using several event catalogues. The CME survey is based on STEREO/EUVI/COR data and the LASCO/CME catalogue$\footnote{http://cdaw.gsfc.nasa.gov/CME\_list/}$. The flare survey is based on the GOES catalogue$\footnote{ftp://ftp.ngdc.noaa.gov/STP/space-weather/solar-data/solar-features/solar-flares/x-rays/goes/}$. For the type II burst survey we distinguished between coronal and interplanetary type II bursts. The information on coronal type II bursts is extracted from a catalogue which collects information from a network of 21 ground-based stations $\footnote{ftp://ftp.ngdc.noaa.gov/STP/space-weather/solar-data/solar-features/solar-radio/radio-bursts/reports/spectral-listings/Type\_II/}$.
The information on interplanetary type II bursts is extracted from the SWAVES (STEREO/WAVES$\footnote{http://ssed.gsfc.nasa.gov/waves/data\_products.html}$ onboard STEREO; \opencite{bougeret08}) catalogue.

\section{Methods}\label{sec:analysis}
For all events under study the EUVI filtergrams were reduced using the SECCHI$\_$PREP routines
available within SolarSoft. We differentially rotated each data set
to a common reference time. In order to enhance faint signatures, we
derived base-ratio (BR) images by dividing each direct image by a
pre-event image. Thus, we derive for
each pixel of the image the relative change of intensity with
respect to the pre-event state. Finally, a median filter was applied
to the images to remove small scale variations.

The kinematical analysis of large-scale disturbances is often
based on visual tracking of the outer edge of the wave front. An
alternative is the semi-automated perturbation profile method
\citep{warmuth04a,podladchikova05,muhr10,long11}, which we also use in this study.
The advantages of the perturbation profile method are that it gives
insight into important physical parameters of EUV waves, \textit{e.g.} amplitude, and that it is observer independent \citep{muhr11}. Nevertheless there are also disadvantages, like the usage of BR images as they highly rely on the preevent images. The problem of BR images is that the base level varies with the spatial position and thus, each position is normalized by a different value. We try to minimize this effect by using a preevent image of the quiet solar corona. On the one hand quantities in physics are defined as the net change and thus, base difference images (BD) would better account for this purpose. On the other hand the usage of BD images leads to similar problems as mentioned above using BR images and additionally, from BR images the relative change (in \%) can be used for further characteristics, \textit{e.g.} the Mach number of the event can be determined \citep{kienreich11,muhr11}. In this sense, neither BR nor BD images may be better than the other, and either of them is simply a proxy of the perturbation that we observe. Thus this algorithm has its limitations.

The perturbation profile algorithm needs some manual input \citep{muhr11}: a) the center of the eruption, and b) the propagation direction under study (which can be chosen by the observer). The center was determined by drawing the earliest observed EUV wave front, transforming the heliocentric coordinates to heliographic coordinates and applying a circular fit to the measurements. The center of this circle is then re-transformed to heliocentric coordinates. We calculate the kinematics of the wave fronts along great circles on the solar surface, studying the propagation in a sector of 45$\degr$ width, in which the wave is most pronounced. Perturbation profiles are obtained by deriving the mean values of all pixels between two constantly growing concentric circles defining annuli with a radial width of 1$\degr$ within the selected propagation sector.

\begin{figure*}
    \centering%
    \includegraphics[width=0.90\textwidth,keepaspectratio=true]{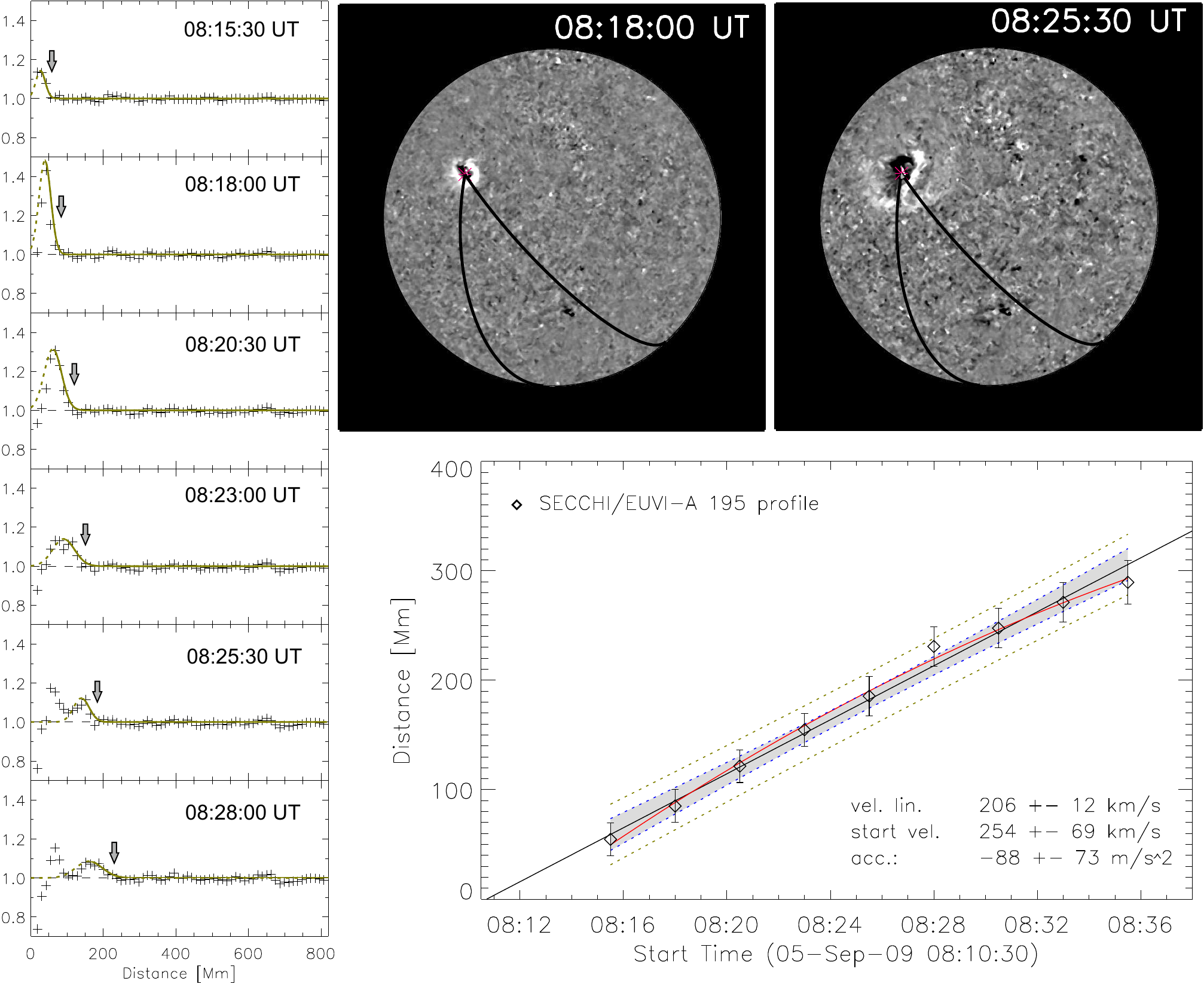}
    \caption[.]{Right, top: Sequence of STEREO/EUVI-A 195$\hbox{\AA}$ BR images for the 5 September 2009 event.
    The black lines indicate great circles through the wave center that determine the propagation direction on which we focus for the wave
    analysis. The FoV is $x=[-1200'', +1200'']$, $y=[-1200'', +1200'']$, with the origin at the center of the Sun. Left: Evolution of perturbation profiles as observed via ST-A for the analyzed sector
    (290$\degr\pm$22.5$\degr$). The crosses are the data points, the solid lines are the fitted Gaussian envelopes to the leading edge of the wave bump while the dashed lines are the mirrored part of the fitted Gaussian envelopes. The arrows indicate the position of the wave fronts.
    Right, bottom: Kinematics of the wave fronts derived via perturbation profiles.
    The black and red solid
    lines indicate the linear and the quadratic least-square fits to the data set, respectively. The grey area within the inner dashed lines indicates the 68\% confidence interval of the linear fit, the area within the outer dashed lines indicate the 95\% confidence interval.} \label{img:20090905A_neu}
\end{figure*}

\begin{figure*}
    \centering%
    \includegraphics[width=0.90\textwidth,keepaspectratio=true]{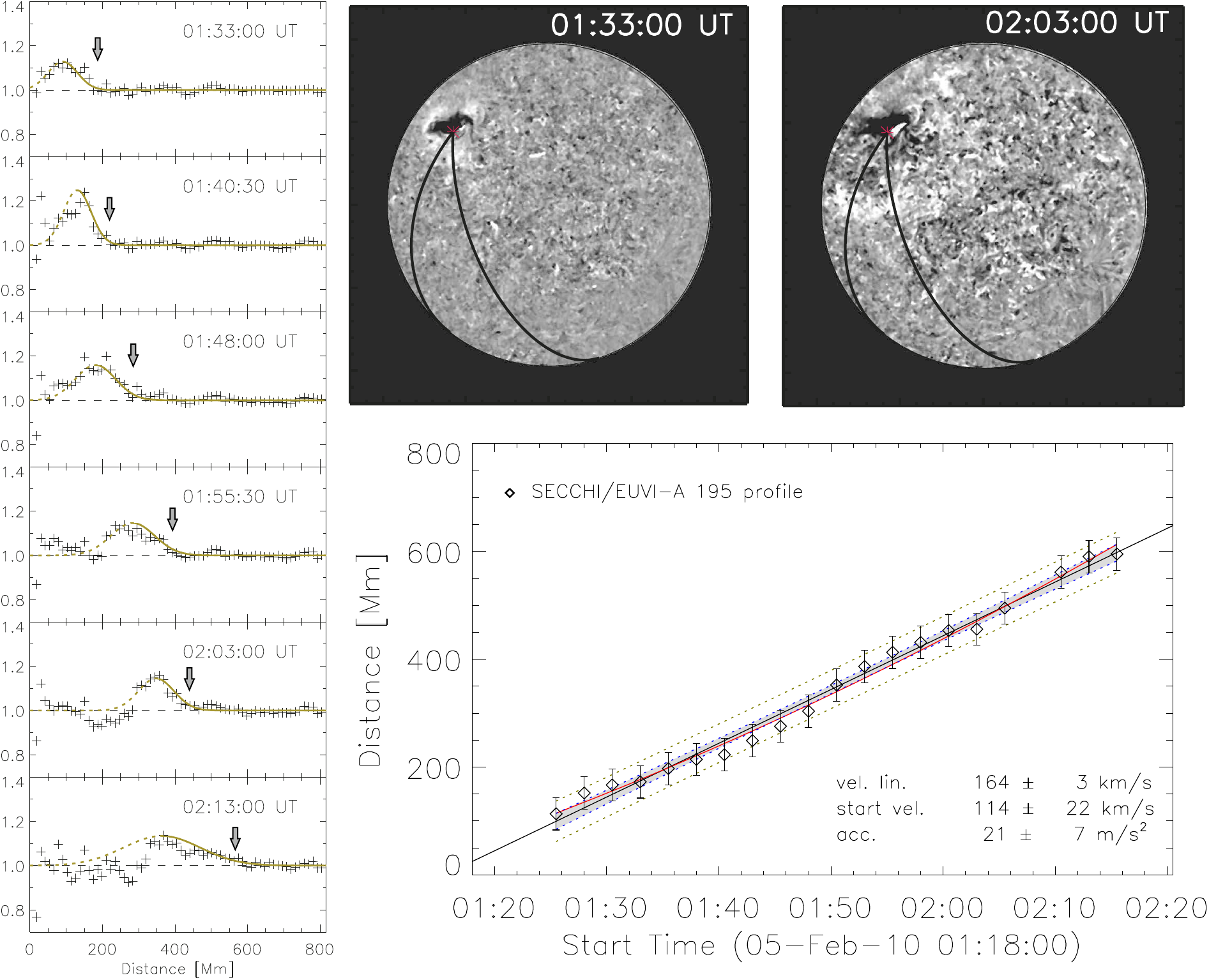}
    \caption[.]{Same as in Figure~\ref{img:20090905A_neu} but for the 5 February 2010 event observed by ST-A.} \label{img:20100205A_b}
\end{figure*}

\begin{figure*}
    \centering%
    \includegraphics[width=0.90\textwidth,keepaspectratio=true]{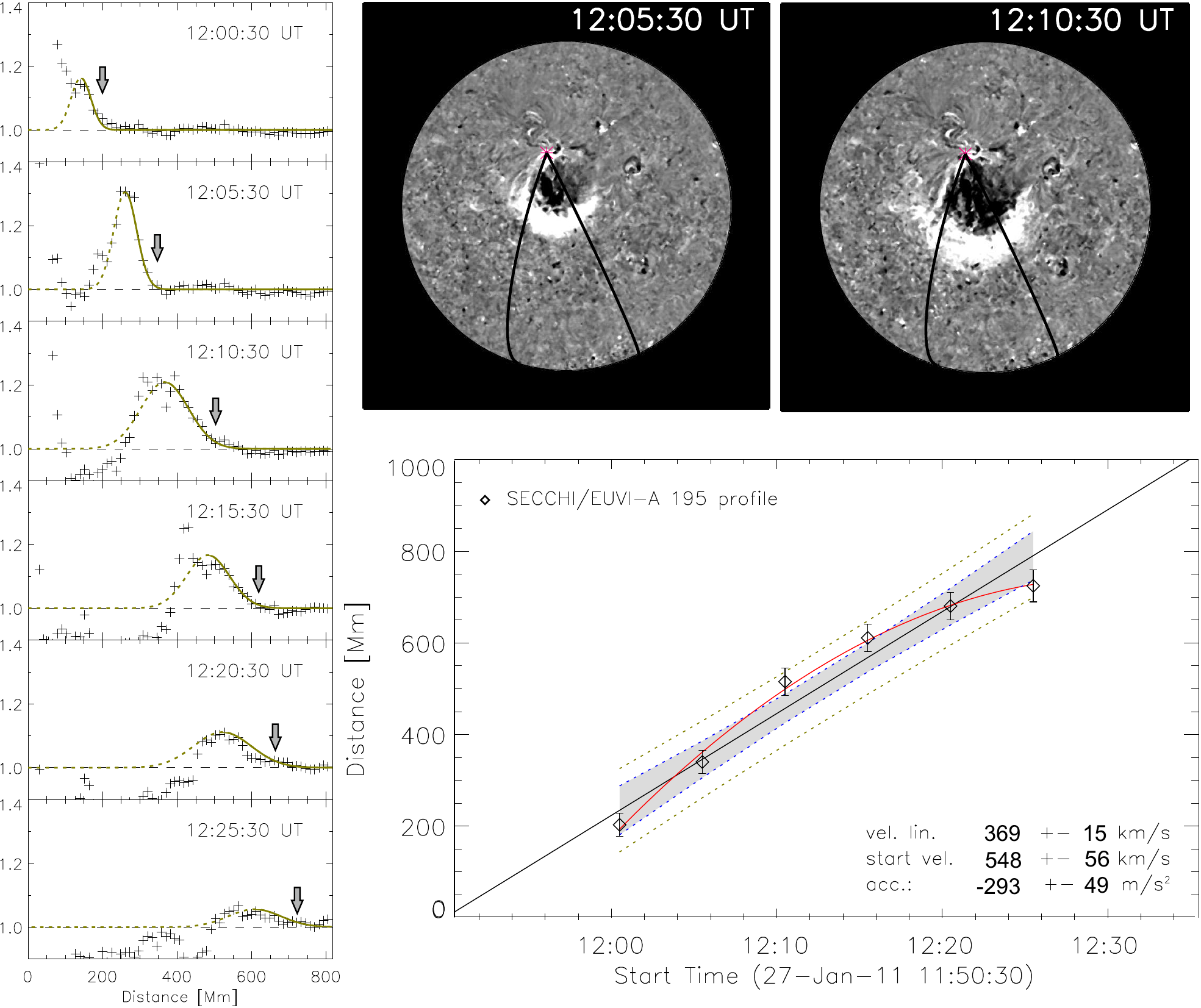}
    \caption[.]{Same as in Figure~\ref{img:20090905A_neu} but for the 27 January 2011 event observed by ST-A.} \label{img:20110127A_b_neu}
\end{figure*}

\begin{figure*}
    \centering%
    \includegraphics[width=0.90\textwidth,keepaspectratio=true]{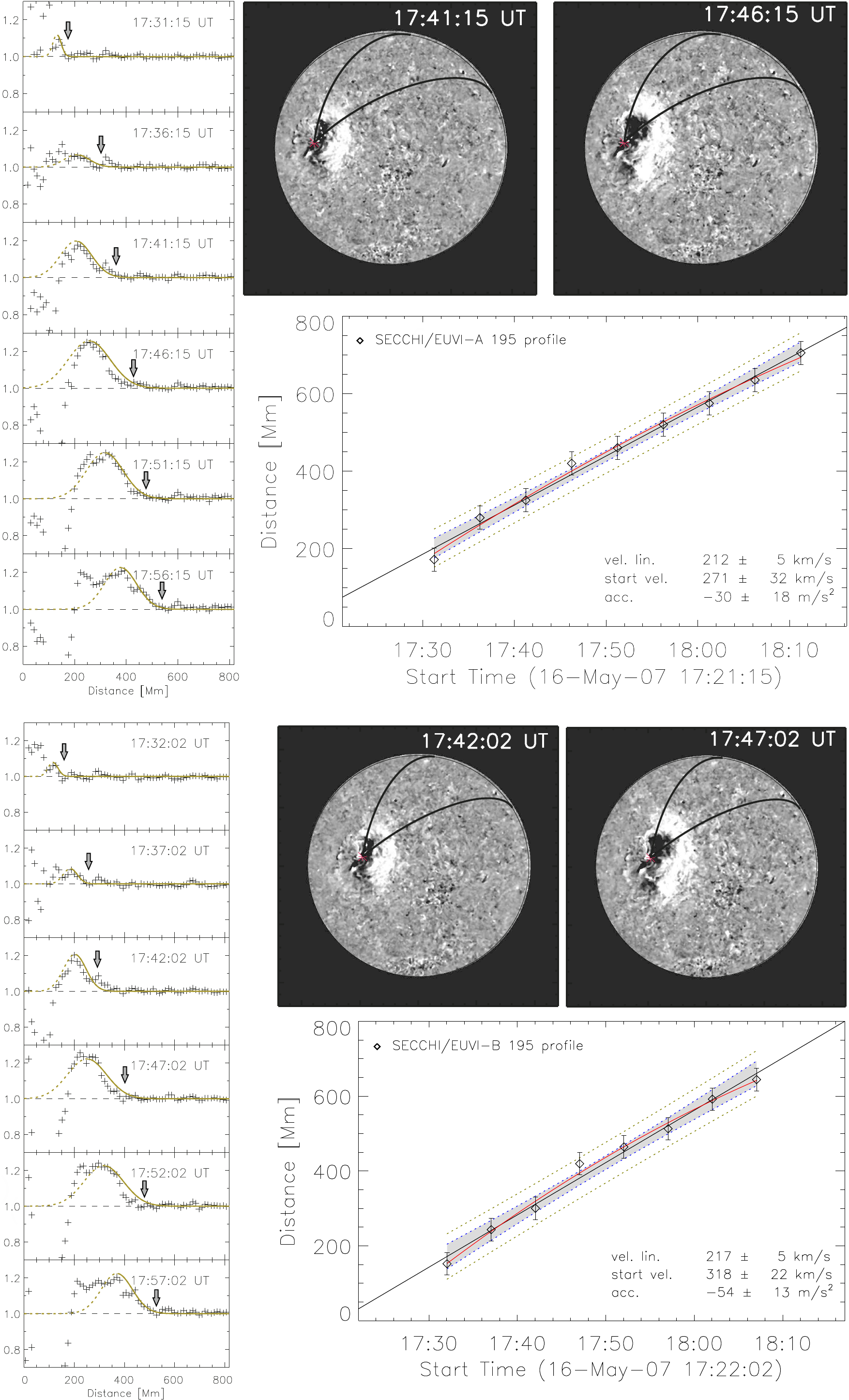}
    \caption[.]{Same as in Figure~\ref{img:20090905A_neu} event observed from ST-A (top) and ST-B (bottom).} \label{img:20070516A}
\end{figure*}

\begin{figure*}
    \centering%
    \includegraphics[width=0.90\textwidth,keepaspectratio=true]{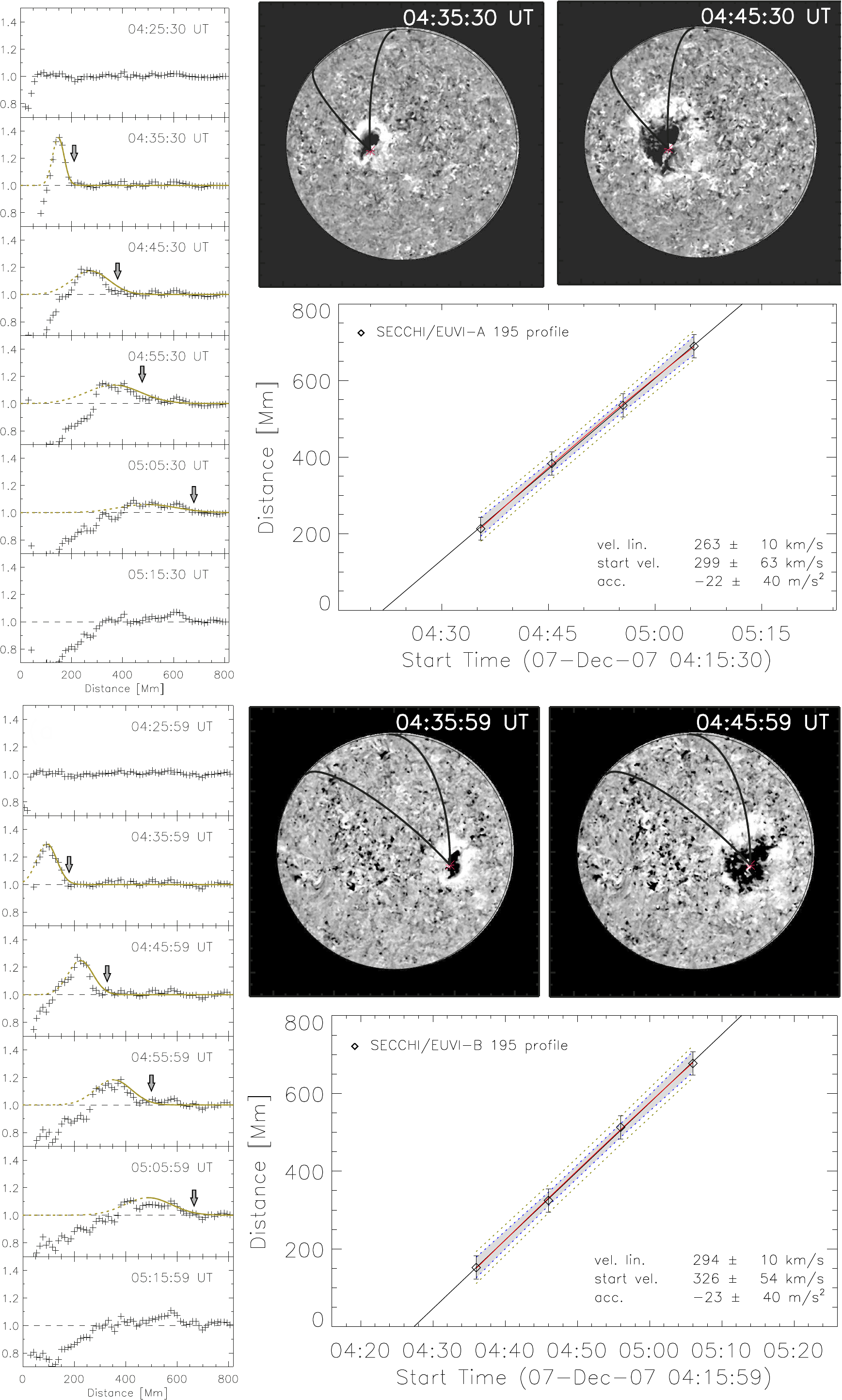}
    \caption[.]{Top: Same as in Figure~\ref{img:20090905A_neu} but for the 7 September 2007 event as observed by ST-A. Bottom: Same as in Figure~\ref{img:20090905A_neu} but for the 7 September 2007 event as observed by ST-B.} \label{img:20071207B}
\end{figure*}

In the perturbation profiles, propagating disturbances can be clearly identified as distinct bumps above the background level (which is 1.0 in base ratio images). Modified Gaussian envelopes emphasizing the leading part of the wave bump are then fitted to the perturbation profiles. Therefore a simple Gaussian fit is applied to the first half of the wave bump from the maximum until the leading edge (shown as bold lines in Figures~\ref{img:20090905A_neu}-\ref{img:20110127A_b_neu}, \ref{img:20070516A}, and \ref{img:20071207B}). The trailing part of the wave bump is not used for the Gaussian envelope fitting process as it may be affected by, \textit{e.g.} enhanced emission from CME flanks, solar flares, stationary brightenings or coronal dimmings behind the EUV wave. For presentation purposes, the fitted Gaussian envelope is mirrored at the position of the peak amplitude (shown as dashed lines in Figures~\ref{img:20090905A_neu}-\ref{img:20110127A_b_neu}, \ref{img:20070516A}, and \ref{img:20071207B}). From the Gaussian fits, we extract the peak amplitude of the disturbance as well as the position of the wave front, applying a threshold level of 2\% above the background.

For each event under study, we derive time-distance plots of the EUV wave front. Linear as well as quadratic least-square
fits are applied to the data points. From the linear fits the mean (constant) propagation
velocities ($v_{\rm{lin}}$) are determined, while the quadratic fits are
used to obtain the `start' velocity ($v_{\rm{start}}$) and the
acceleration/deceleration of the EUV wave during its propagation. The time of first wave
observation is taken as start time. The velocity derived from the quadratic fit at that instant is used as an estimate of the `start velocity' of the EUV wave. The fits and the errors on the fitting results are obtained with the bootstrapping method \citep{efron79}. To this aim, we added random errors (with a maximum of $\pm 35$~Mm) to the position measurements, fitted the data set, and repeated this procedure a thousand times. The mean values and the standard deviations are used as the resulting kinematical fit parameters and related errors.

In addition, we gain important information on the wave events by extracting characteristic parameters from the perturbation profiles, \textit{e.g.} the
perturbation amplitude evolution as well as the peak amplitude
$A_{\rm{max}}$. We also derive the build-up time $\Delta t_{\rm{build}}$ that is
calculated as the time difference between the first observation and the time of the peak amplitude.
The impulsiveness $I$ of an event is calculated by
dividing the peak amplitude by its build up time, $I=A_{\rm{max}}/\Delta t_{\rm{build}}$.

\section{Results}\label{sec:results}
\subsection{Statistical Analysis of EUV Wave Kinematics}\label{sec:results_statistics}
An overview of all events under study is given in
Table~\ref{tab:table1}~--~\ref{tab:table4}. For each EUV wave event we list the following information:
(1) observation date and start time; (2) observing spacecraft; (3) mean velocity $v_{\rm{lin}}$; (4) start velocity $v_{\rm{start}}$; (5) acceleration $a$;
(6) peak amplitude $A_{\rm{max}}$; (7) CME association and CME velocity; (8) SXR flare association and classification; (9) coronal and/or interplanetary type II radio burst
association.

\begin{sidewaystable*}
\centering
\begin{tabular}{l cccccclc}
  \hline\hline
     Event & S/C & \multicolumn{1}{c}{\textit{v}$_{\rm{lin}}$ }  & \multicolumn{1}{c}{\textit{v}$_{\rm{start}}$}  & \multicolumn{1}{c}{$a$ } & \textit{A}$_{\rm{max}}$ & CME \textit{v}& Flare & Type II\\
     & & [km~s$^{-1}$]& [km~s$^{-1}$] & [m~s$^{-2}$] &&[km~s$^{-1}$] &&\\
     \hline
\textbf{2007 May 16 17:25}$^{\dag\dag}$&A& 212$\pm$5 & 271$\pm$32 & $-30\pm18$ & 1.26 &371&C29 1719 1753 1741&c;--- \\
  \textbf{2007 May 16 17:25}$^{\dag\dag}$&B& 217$\pm$5 & 318$\pm$22 & $-54\pm13$ & 1.26 &371&C29 1719 1753 1741&c;--- \\
  \textbf{2007 May 19 12:40}$^{\dag\dag}$&A& 338$\pm$13 & 373$\pm$58 & $-85\pm131$ & 1.61 &958&B95 1248 1319 1302&c;i \\
 \textbf{2007 May 19 12:40}$^{\dag\dag}$&B& 354$\pm$17 & 397$\pm$31 & $-79\pm101$ & 1.61 &958&B95 1248 1319 1302&c;i\\
  \textbf{2007 May 22 14:20}$^{\dag\dag}$&A& 242$\pm$8 & 254$\pm$28 & $-20\pm29$ & 1.21 &544&B39 1430 1519 1447&---;---\\
  \textbf{2007 May 22 14:20}$^{\dag\dag}$&B& 277$\pm$8 & 287$\pm$28 & $-16\pm29$ & 1.20 &544&B39 1430 1519 1447&---;---\\
  2007 May 23 07:10&B& 239$^{\dag\dag}$$\pm$7 & 252$\pm$28 & $-22\pm21$ & 1.38 &679&B53 0715 0750 0732&---;---\\
\textbf{2007 Dec 07 04:25}$^{\dag\dag}$&A& 263$\pm$10 & 299$\pm$61 &$-22\pm39$ & 1.37 &284&B14 0435 0455 0441&---;---\\
  \textbf{2007 Dec 07 04:25}$^{\dag\dag}$&B& 294$\pm$10 & 326$\pm$54 &$-23\pm40$ & 1.30 &284&B14 0435 0455 0441&---;---\\
  2008 Jan 07 01:45$^{\dag\dag}$&B& 277$\pm$22 & 289$\pm$93 &$-40\pm66$ & 1.49 &282&B12 0226 0245 0234&---;---\\
  2008 Mar 25 18:45$^{\dag\dag}$&B& 463$\pm$16 & 520$\pm$90 &$-136\pm87$ & 2.60 &1103&M17 1836 1913 1856&---;i\\
  2008 Apr 16 19:25$^{\dag\dag}$&A& 243$\pm$22 & 309$\pm$47 & $-162\pm157$ & 1.18 &243&A52 &---;---\\
\textbf{2008 Apr 26 13:35}$^{\dag\dag}$&A& 203$\pm$32 & 311$\pm$28 & $-186\pm171$ & 1.23 &515&B38 1354 1438 1408&c;---\\
  \textbf{2008 Apr 26 13:35}$^{\dag\dag}$&B& 236$\pm$11 & 293$\pm$29 & $-101\pm171$ & 1.56 &515&B38 1354 1438 1408&c;---\\
\textbf{2008 May 16 17:55}&A& 148$\pm$21 & 236$\pm$104 &$-201\pm163$ & 1.34 &---&---&---;---\\
  \textbf{2008 May 16 17:55}&B& 148$\pm$21 & 235$\pm$110 &$-199\pm163$ & 1.29 &---&---&---;---\\
  2008 Jul 20 12:55&B& 194$\pm$10 & 227$\pm$103 &$-57\pm41$ & 1.43 &yes&---&---;---\\
  \hline
\end{tabular}
\caption{EUV wave events under study.
We list the start time of the event, the observing spacecraft (ST-A, ST-B), the mean velocity \textit{v}$_{\rm{lin}}$,
start velocity \textit{v}$_{\rm{start}}$ as well as the acceleration $a$ of both methods and the peak perturbation amplitude values \textit{A}$_{\rm{max}}$ extracted from the perturbation profiles. The association with CME, flare, and type II burst (i: interplanetary; c: coronal) observations is also shown. The shortcut n/a means no available data for the specific spacecraft. The `yes' means a feature is detectable but no further measurement was possible. The minus (-) means no feature detected. Events in bold face are detectable in both ST-A and ST-B. The events marked with a $^{\dag\dag}$ are also measured in SDO/AIA by Nitta \textit{et al.} (2014).}             
\label{tab:table1}      
\end{sidewaystable*}

\begin{sidewaystable*}
\centering
\begin{tabular}{l cccccccc}
  \hline\hline
Event & S/C & \multicolumn{1}{c}{\textit{v}$_{\rm{lin}}$ }  & \multicolumn{1}{c}{\textit{v}$_{\rm{start}}$}  & \multicolumn{1}{c}{$a$ } & \textit{A}$_{\rm{max}}$ & CME \textit{v}& Flare & Type II\\
     & & [km~s$^{-1}$]& [km~s$^{-1}$] & [m~s$^{-2}$] &&[km~s$^{-1}$] &&\\
     \hline
  \textbf{2008 Jul 21 08:05}&A& 195$\pm$20 & 375$\pm$75 &$-89\pm39$ & 1.22 &---&---&---;---\\
  \textbf{2008 Jul 21 08:05}&B& 267$\pm$7 & 357$\pm$109 &$-161\pm21$ & 1.51 &---&---&---;---\\
  2008 Oct 01 06:45&A& 210$\pm$17 & 356$\pm$68 &$-251\pm97$ & 1.27 &158&---&---;---\\
  2009 Feb 10 23:05$^{\dag\dag}$&B& 200$\pm$17 & 201$\pm$61 &$-1\pm90$ & 1.43 &273&B13 2300 2342 2311&---;---\\
  2009 Feb 12 15:55$^{\dag\dag}$&B& 260$\pm$11 & 439$\pm$41 & $-325\pm40$ & 1.54 &356&B41 1610 1631 1619&---;---\\
  2009 Feb 13 05:35$^{\dag\dag}$&B& 221$\pm$25 & 231$\pm$72 & $-23\pm40$ & 1.39 &440&B23 0535 0604 0547&---;---\\
  \textbf{2009 Feb 27 07:05}$^{\dag\dag}$&A& 187$\pm$14 & 238$\pm$46 &$-61\pm39$ & 1.43 &222&A30 &---;---\\
  \textbf{2009 Feb 27 07:05}$^{\dag\dag}$&B& 190$\pm$14 & 241$\pm$66 &$-60\pm22$ & 1.31 &222&A30 &---;---\\
  2009 Apr 26 13:25&A& 179$\pm$5 & 203$\pm$50 & $-25\pm13$ & 1.20 & --- &---& n/a; ---\\
  2009 Sep 05 08:15&A& 206$\pm$12 & 254$\pm$69 & $-88\pm73$ & 1.48 &171&---&---;---\\
  2009 Oct 12 05:15&A& 158$\pm$10 & 265$\pm$39 &$-152\pm37$ & 1.32 &yes&---& ---;---\\
  2009 Oct 26 11:35&B& 170$\pm$7 & 100$\pm$96 & $+47\pm24$ & 1.48 &yes&B17 1144 1209 1152& ---;---\\
  2009 Nov 15 11:35&A& 161$\pm$9 & 137$\pm$104 &$+23\pm34$ & 1.39 &407&---&n/a;--- \\
  2009 Dec 05 10:05&B& 228$\pm$13 & 239$\pm$65 &$-118\pm78$ & 1.29 &371&---&n/a;--- \\
  2009 Dec 22 04:55$^{\dag\dag}$&A& 417$\pm$22 & 514$\pm$95 &$-207\pm161$ & 2.20 &318&C72 0450 0500 0456&c;
  --- \\
  2009 Dec 23 10:15$^{\dag\dag}$&A& 340$\pm$11 & 415$\pm$43 &$-167\pm57$ & 1.90 &358&C64 1009 1020 1017&---;---\\
  2010 Jan 17 03:55&B& 277$\pm$17 & 278$\pm$21 & $-1\pm8$  & 1.59 &350&---&n/a;i \\
  \hline
\end{tabular}
\caption{Same as in Table~1.}             
\label{tab:table2}      
\end{sidewaystable*}

\begin{sidewaystable*}
\centering
\begin{tabular}{l cccccccclc}
  \hline\hline
 Event & S/C & \multicolumn{1}{c}{\textit{v}$_{\rm{lin}}$ }  & \multicolumn{1}{c}{\textit{v}$_{\rm{start}}$}  & \multicolumn{1}{c}{$a$ } & \textit{A}$_{\rm{max}}$ & CME \textit{v}& Flare & Type II\\
     & & [km~s$^{-1}$]& [km~s$^{-1}$] & [m~s$^{-2}$] &&[km~s$^{-1}$] &&\\
     \hline
  2010 Jan 24 20:25&A& 197$\pm$10 & 218$\pm$72 & $-25\pm56$ & 1.14 &yes&B20 2037 2105 2051&---;--- \\
  2010 Jan 26 17:05&A& 205$\pm$11 & 361$\pm$48 & $-226\pm56$ & 1.18  &228&B32 1701 1708 1705&---;---\\
  2010 Jan 26 18:35&A& 273$\pm$19 & 304$\pm$138 &$-145\pm135$ & 1.16 &300&B67 1740 1758 1751&---;---\\
  2010 Jan 30 14:05&A& 235$\pm$9 & 225$\pm$58 &$-67\pm45$ & 1.59 &219&n/a&n/a;--- \\
  2010 Feb 05 01:25&A& 164$\pm$3 &  114$\pm$22 & $+21\pm7$ & 1.28 &328&---&---;---\\
  2010 Feb 10 16:05&B& 213$\pm$11 & 211$\pm$65 & $-4\pm51$ & 1.24 &538&---&---;---\\
  2010 Feb 18 18:45&A& 220$\pm$13 & 214$\pm$84 &$-9\pm72$ & 1.29  &223&---&---;--- \\
  2010 Mar 04 13:35&A& 258$\pm$12 & 362$\pm$81 & $-143\pm73$ & 1.39 &374&B60 1328 1333 1331&c;---\\
  2010 Mar 04 23:15&A& 188$\pm$7 & 186$\pm$77 & $-7\pm29$ & 1.29 &277&B23 2329 0040 2353&---;---\\
  2010 Mar 13 23:35&B& 225$\pm$6 & 412$\pm$23 & $-173\pm18$ & 1.13 &351&C15 2335 0023 2349&---;i\\
  2010 Mar 14 07:15&B& 230$\pm$15 & 278$\pm$84 &$-80\pm84$ & 1.69 &208&B24 0745 0759 0754&---;---\\
  2010 Mar 25 01:15&B& 375$\pm$22 & 466$\pm$109 &$-211\pm157$ & 1.45 &338&n/a&n/a;---\\
  2010 Apr 28 22:35&B& 220$\pm$28 & 230$\pm$141 &$-18\pm83$ & 1.19 &yes&n/a&n/a;---\\
  2010 Apr 29 00:25&B& 233$\pm$12 & 305$\pm$59 &$-96\pm49$ & 1.14 &138&n/a&n/a;---\\
  2010 Apr 29 02:55&B& 225$\pm$10 & 305$\pm$58 &$-52\pm58$ & 1.17  &178&n/a&n/a;---\\
  2010 Apr 29 06:25&B& 321$\pm$13 & 350$\pm$39 & $-52\pm41$ & 1.22 &187&n/a&n/a;---\\
  2010 Apr 30 23:35&B& 168$\pm$11 & 440$\pm$46 &$-115\pm53$ & 1.28  &yes&---&---;---\\
   \hline
\end{tabular}
\caption{Same as in Table~1.}             
\label{tab:table3}      
\end{sidewaystable*}

\begin{sidewaystable*}
\centering
\begin{tabular}{l cccccccccc}
  \hline\hline
Event & S/C & \multicolumn{1}{c}{\textit{v}$_{\rm{lin}}$ }  & \multicolumn{1}{c}{\textit{v}$_{\rm{start}}$}  & \multicolumn{1}{c}{$a$ } & \textit{A}$_{\rm{max}}$ & CME \textit{v}& Flare & Type II\\
     & & [km~s$^{-1}$]& [km~s$^{-1}$] & [m~s$^{-2}$] &&[km~s$^{-1}$] &&\\
     \hline
  2010 Jun 12 00:55$^{\dag}$&A& 387$\pm$15 & 601$\pm$63 &$-331\pm84$ & 1.21 &486&M20 0030 0102 0057&c;i\\
  2010 Jun 12 04:05&A& 342$\pm$11 & 477$\pm$82 &$-188\pm53$ & 1.09 &303&C10 0357 0417 0406&---;---\\
  2010 Jun 12 09:15$^{\dag}$&A& 251$\pm$10 & 266$\pm$46 & $-97\pm35$ & 1.12 &382&C61 0902 0922 0917&c;---\\
  2010 Jun 13 05:35$^{\dag}$&A& 316$\pm$9 & 321$\pm$47 &$-8\pm43$ & 1.49 &349&M10 0530 0544 0539&c;---\\
  2010 Jun 16 03:05&A& 251$\pm$14 & 371$\pm$69 &$-198\pm87$ & 1.24 &275&---&---;---\\
  2010 Jul 06 10:05&A& 161$\pm$11 & 165$\pm$43 & $-4\pm53$ & 1.59 &369&B10 1002 1016 1011&---;--- \\
  2010 Jul 31 05:25&B& 245$\pm$15 & 254$\pm$86 & $-16\pm90$ & 1.11 &yes&B48 0515 0532 0529&---;--- \\
  2010 Aug 07 18:15$^{\dag}$&B& 423$\pm$15 & 709$\pm$57 &$-355\pm83$ & 1.64 &400&M10 1755 1847 1824&---;i\\
  2010 Aug 14 09:05$^{\dag}$&A& 339$\pm$9 & 353$\pm$44 &$-21\pm34$ & 1.49  &657&C44 0938 1031 1005&c;--- \\
  2010 Aug 18 04:45&A& 215$\pm$9 & 300$\pm$72 &$-81\pm35$ & 1.26 &650&C45 0445 0651 0548&---;i\\
  2010 Sep 07 14:15&B& 365$\pm$8 & 368$\pm$41 &$-4\pm36$ & 1.16 &221&---&---;---\\
  2010 Sep 08 22:55&A& 323$\pm$14 & 381$\pm$76 & $-19\pm22$ & 1.52 &818&C33 2305 0010 2333&---;i\\
  2010 Sep 14 06:55&B& 169$\pm$7 & 245$\pm$22 & $-76\pm19$ & 1.33 &380&--- &---;---\\
  2011 Jan 27 08:45&A& 341$\pm$13 & 610$\pm$71 &$-340\pm48$ & 1.22 &455&B66 0840 0853 0850&c;---\\
  2011 Jan 27 12:00$^{\dag}$&A& 369$\pm$15 & 548$\pm$56 &$-293\pm49$ & 1.30 &413&B19 1213 1226 1218&c;i\\
  2011 Jan 27 19:55&A& 317$\pm$11 & 345$\pm$61 &$-31\pm32$ & 1.40 &416&C15 1950 2019 2003&---;---\\
  \hline
\end{tabular}
\caption{Same as in Table~1. The events marked with a $^{\dag}$ are also measured in SDO/AIA by Nitta \textit{et al.} 2013.}             
\label{tab:table4}      
\end{sidewaystable*}

\begin{figure*}
    \centering%
    \includegraphics[width=0.8\textwidth,keepaspectratio=true]{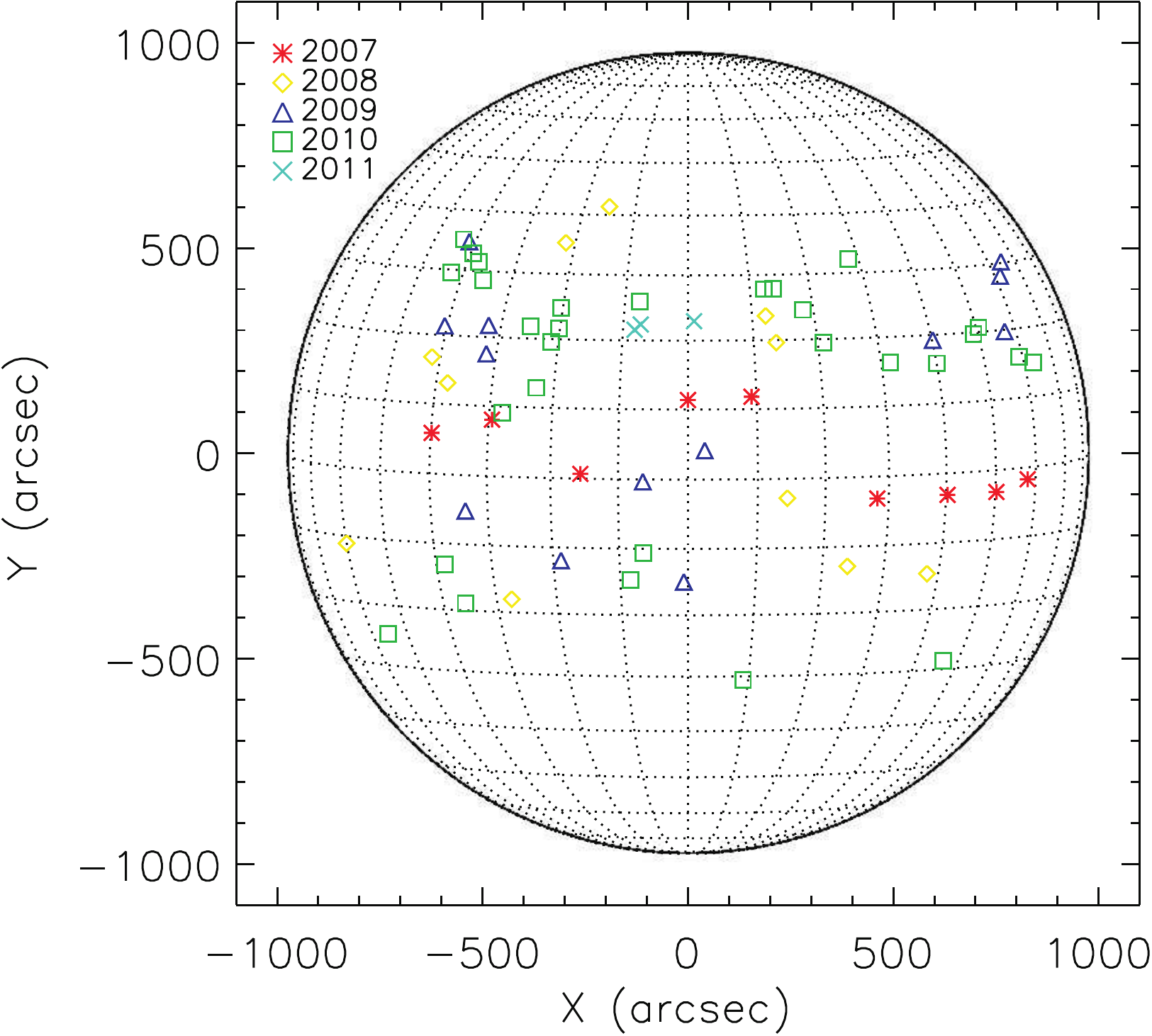}
    \caption[.]{Positions of the EUV wave events source regions on the solar disk as observed from the actual ST-A or ST-B vantage points. Different colors and symbols represent the different years of occurrence.} \label{img:map_start_location_b}
\end{figure*}

Figure~\ref{img:map_start_location_b} shows the source center of the EUV waves under study, in the view of the STEREO s/c at the time of observations. The figure also reveals the transition from solar cycle 23 to 24, with the change of the eruption centers from equatorial locations at the end of solar cycle 23 to higher latitudes in the beginning of solar cycle 24.

The kinematical analysis via the perturbation profile method is
applied to all 60 EUV wave events under study.
In Figures~\ref{img:20090905A_neu} -- \ref{img:20110127A_b_neu} we show three examples for the analysis conducted from our data set. The events occurred on 5 September 2009, 5 February 2010 and 27 January 2011. Each figure shows two base ratio images of the event (upper right), the evolution of the perturbation profiles together with the Gaussian fits (left) and the EUV wave kinematics derived together with the linear and quadratic fits (bottom).

For all wave events under study we observe a clear steepening of the perturbation
amplitude reaching the peak amplitude
$A_{\rm{max}}$ 5 up to 30 min after the first wave
signature is observed. The mean value for the build-up time $t_{\rm{build}}$ for all events under study is determined to be $\approx$ 9~min. After the peak amplitude has been reached the wave pulses decay and broaden until the transient can no longer be traced. The perturbation
amplitudes of all events under study follow a characteristic
evolution showing a steepening in the beginning, and a decaying and broadening in
the end (see left panels in Figures~\ref{img:20090905A_neu} -- \ref{img:20110127A_b_neu}).

\begin{figure*}
    \centering%
    \includegraphics[width=1.0\textwidth,keepaspectratio=true]{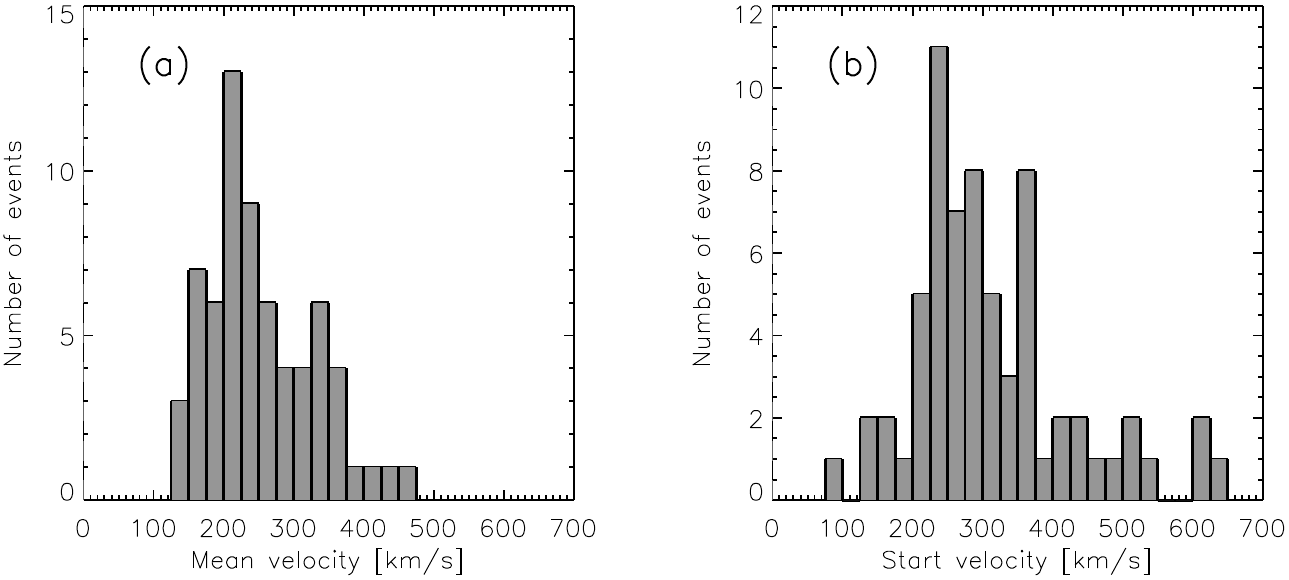}
    \caption[.]{Distributions of the mean (a) and start (b) velocities of the EUV wave events under study.} \label{img:histogram}
\end{figure*}

In Tables~\ref{tab:table1} to \ref{tab:table4}, we list the kinematical results for the whole EUV event sample. Figure~\ref{img:histogram}(a) shows the distribution of the mean
velocities lying in a range of $130-470$~km~s$^{-1}$ with an arithmetic mean of $254\pm76$~km~s$^{-1}$ and a median of
$239\pm73$~km~s$^{-1}$. The distribution of the start velocities derived from the quadratic fits at the time of the first wave front observations are shown in Figure~\ref{img:histogram}(b). The distribution ranges from $100$ to $630$~km~s$^{-1}$. The arithmetic mean is
$312\pm115$~km~s$^{-1}$, the median
$287\pm93$~km~s$^{-1}$. These results are similar to
the outcome of the statistical studies by \inlinecite{thompson09} who report a mean velocity of $215\pm103$~km~s$^{-1}$ based on EIT data
and \inlinecite{warmuth11} who report a mean velocity of
$264\pm138$~km~s$^{-1}$ using a combined
EIT and EUVI data set.

\begin{figure*}
    \centering%
    \includegraphics[width=0.65\textwidth,keepaspectratio=true]{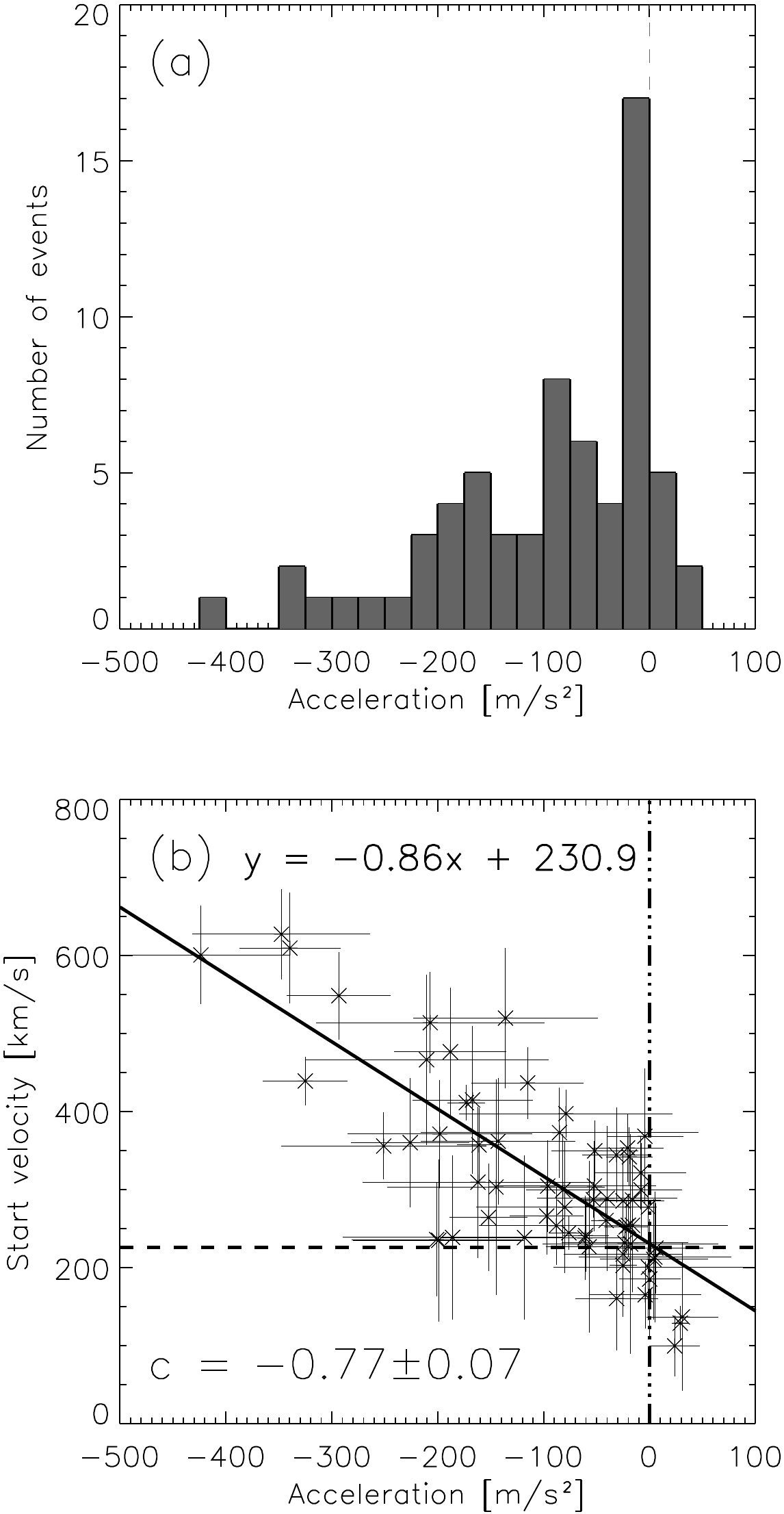}
    \caption[.]{(a) Distribution of the mean acceleration derived from the quadratic fits to the EUV wave kinematics.
    (b) Start velocities $v_{\rm{start}}$ versus mean acceleration $a$. The solid line represents the least-square fit to the data. The dashed dotted line vertical line indicates constant speed ($a=0$~m~s$^{-2}$). The dashed horizontal line is plotted through the intersection of the regression line and the $a=0$~m~s$^{-2}$ line, giving $v\approx230$~km~s$^{-1}$.} \label{img:acceleration}
\end{figure*}

Figure~\ref{img:acceleration}(a) shows the distribution of the accelerations, which are in the range of $a=[-460,50]$~m~s$^{-2}$. The distribution shows that most data points are in the negative range (the arithmetic mean is $-97\pm103$~m~s$^{-2}$), indicating deceleration of the EUV waves during propagation. To quantify the significance, we determined the errors on the acceleration values derived from the quadratic fits using bootstrapping: for the whole event sample, we obtain an arithmetic mean for the errors of 49~m~s$^{-2}$ (median 54~m~s$^{-2}$). Consequently, we attribute all waves which have acceleration values in the range of $-50 \leq a \leq 50$~m~s$^{-2}$ still as consistent with a propagation of constant speed. This group comprises 48\% of the events, whereas 52\% of the EUV waves under study show a significant deceleration.

Figure~\ref{img:acceleration}(b) shows the scatter plot of the start velocity $v_{\rm{start}}$ against the acceleration $a$. The black solid line represents the least-square linear fit to the data set revealing a strong anti-correlation with a correlation coefficient of $c=-0.77\pm0.07$. Thus, the faster
the wave initially, the stronger the deceleration during its propagation. To mark the value of zero acceleration we plotted a vertical dashed-dotted line at this position. The intersection of the least-square linear fit and the zero acceleration line lies at a velocity value of $v_{\rm{fm}}\approx 230$~km~s$^{-1}$ corresponding to the typical fast-mode magnetosonic speed in the quiet solar corona. Thus, events with initial speeds higher than the fast magnetosonic speed decelerate, whereas events with initial speeds close to the magnetosonic speed show constant speed during wave propagation. This relationship provides strong support that EUV waves are fast-mode MHD waves propagating in the solar corona.

\subsection{Characteristics of Perturbation Profiles}~\label{sec:results_correlation}
\begin{figure*}
    \centering%
    \includegraphics[width=1.0\textwidth,keepaspectratio=true]{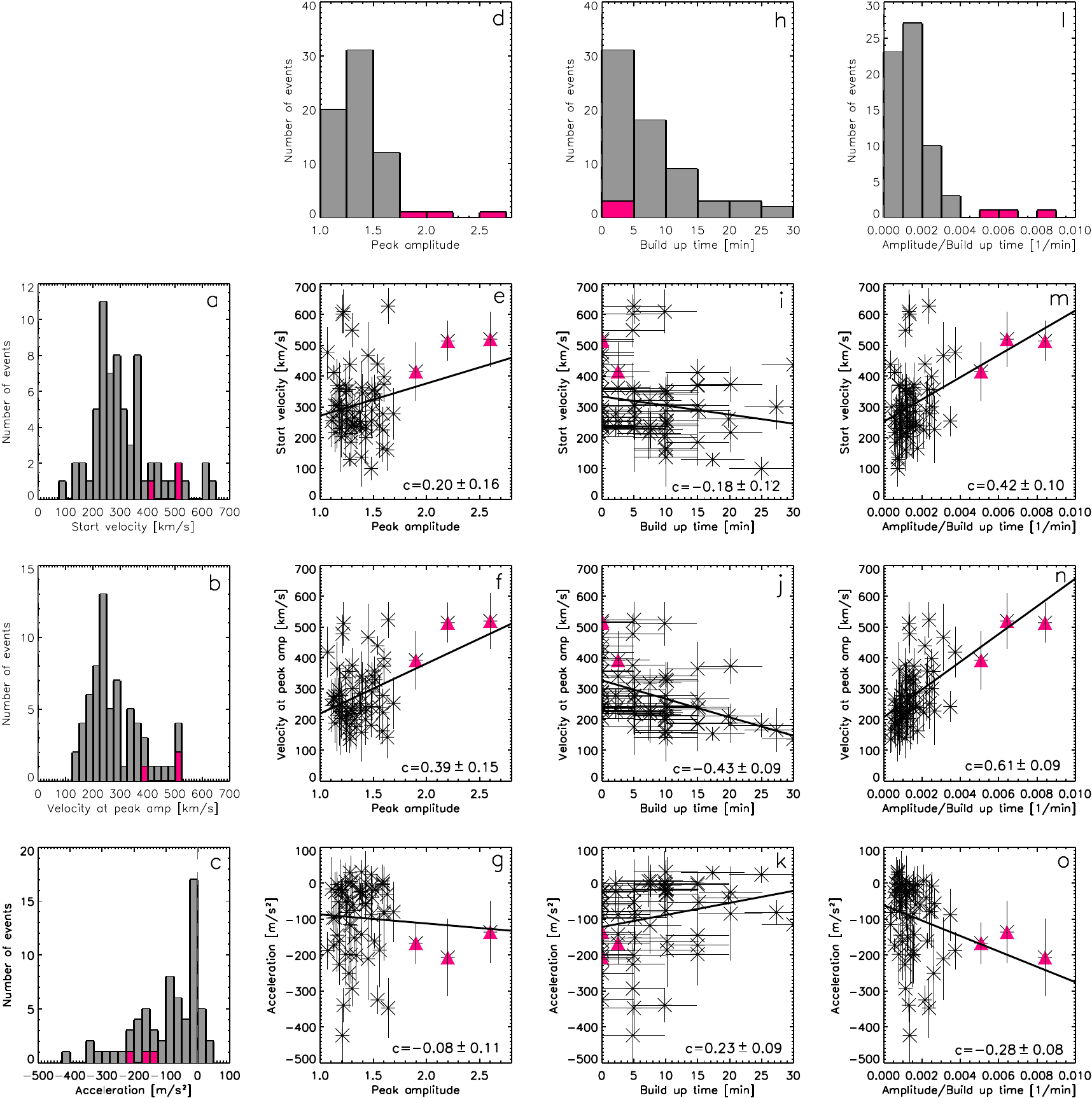}
    \caption[.]{Distributions and scatter plots of EUV wave parameters: start velocity, velocity at peak amplitude, acceleration; peak amplitude, build-up time and impulsiveness of the perturbation profile evolution. The three pink triangles in each panel represent the three data points on which the correlation coefficient of the panels e, f, g, m, n, and o highly rely on.} \label{img:correlation_muhr2011_4mal4_new}
\end{figure*}

Perturbation profiles are used to extract kinematical measurements
and to give information on the perturbation itself, \textit{e.g.} the peak amplitude $A_{\rm{max}}$, the wave pulse evolution, the build-up time $\Delta t_{\rm{build}}$, and the impulsiveness $I$ of the events. In order to identify general characteristics and relationships, which intrinsically describe the evolution of the events, we correlated these parameters. The correlation plots are displayed in Figure~\ref{img:correlation_muhr2011_4mal4_new}.

The highest correlations are obtained for the correlations of the velocity at peak amplitude $v_{\rm{\textit{A}max}}$ against the peak amplitude $A_{\rm{max}}$ (correlation coefficient $c=0.39\pm0.15$), the build-up time $\Delta t_{\rm{build}}$ ($c=-0.43\pm0.09$) and the impulsiveness $I$ ($c=0.61\pm0.09$) of the events. A positive correlation of $v_{\rm{\textit{A}max}}$ against $A_{\rm{max}}$ means that faster events show more pronounced peak amplitudes. The anti-correlation of $v_{\rm{\textit{A}max}}$ against $\Delta t_{\rm{build}}$ is a result of faster events reaching the peak amplitude in shorter time. The positive correlation of $v_{\rm{\textit{A}max}}$ against $I$ implies that faster events are more impulsive as they compress the surrounding plasma within a shorter period. We stress that the correlation is highly depending on three extremely strong and fast wave events which are marked by pink triangles in each panel of Figure~\ref{img:correlation_muhr2011_4mal4_new}. Without these measurements the correlation would be less pronounced. During the considered time span around solar minimum, pronounced events except the three mentioned above are missing. This can also be seen by the rather low mean value of the peak amplitude $\bar{A}_{\rm{max}}=1.39\pm0.25$.

\subsection{Association Rates with CMEs, Flares and Type II Bursts}
The association rates of the EUV waves with CMEs, solar flares, and type II radio
bursts provide additional constraints for identifying
the physical nature of the transients and possible
generation or launching mechanisms.

For the 60 EUV wave events under study we found 57 associated CMEs and thus, an
association rate of $\approx$~95~\% (\textit{cf.} Tables~\ref{tab:table1} -- \ref{tab:table4}). For 50 CMEs the projected mean velocity was listed in the LASCO catalogue. For the other seven events the velocity was not listed in the LASCO catalogue though a CME structure was clearly observable in STEREO imagery.
For the flare association rate we note that in 20\% (11 out of 60 events) of all EUV wave observations there was no available data set from the GOES satellite as the wave event was launched behind the solar limb as seen from Earth. Thus, there is neither a positive nor a negative report about a possible association for these events. For the remaining 49 events we find 36 associated solar flares giving an association rate of $\approx$74\%.
Additionally we checked for associated type II bursts distinguishing between coronal and interplanetary type~II bursts.
Examining the SWAVES catalogue we found 9 associated interplanetary type II radio bursts giving an association rate of 15\%.
Coronal type II bursts are identified in radio spectra from a network of ground-based stations. For $\approx$20\% of the events under study no data are available. For the remaining 49 events we find 11 associated coronal type II bursts which gives an association rate of $\approx$22\%.

\begin{figure*}
    \centering%
    \includegraphics[width=1.0\textwidth,keepaspectratio=true]{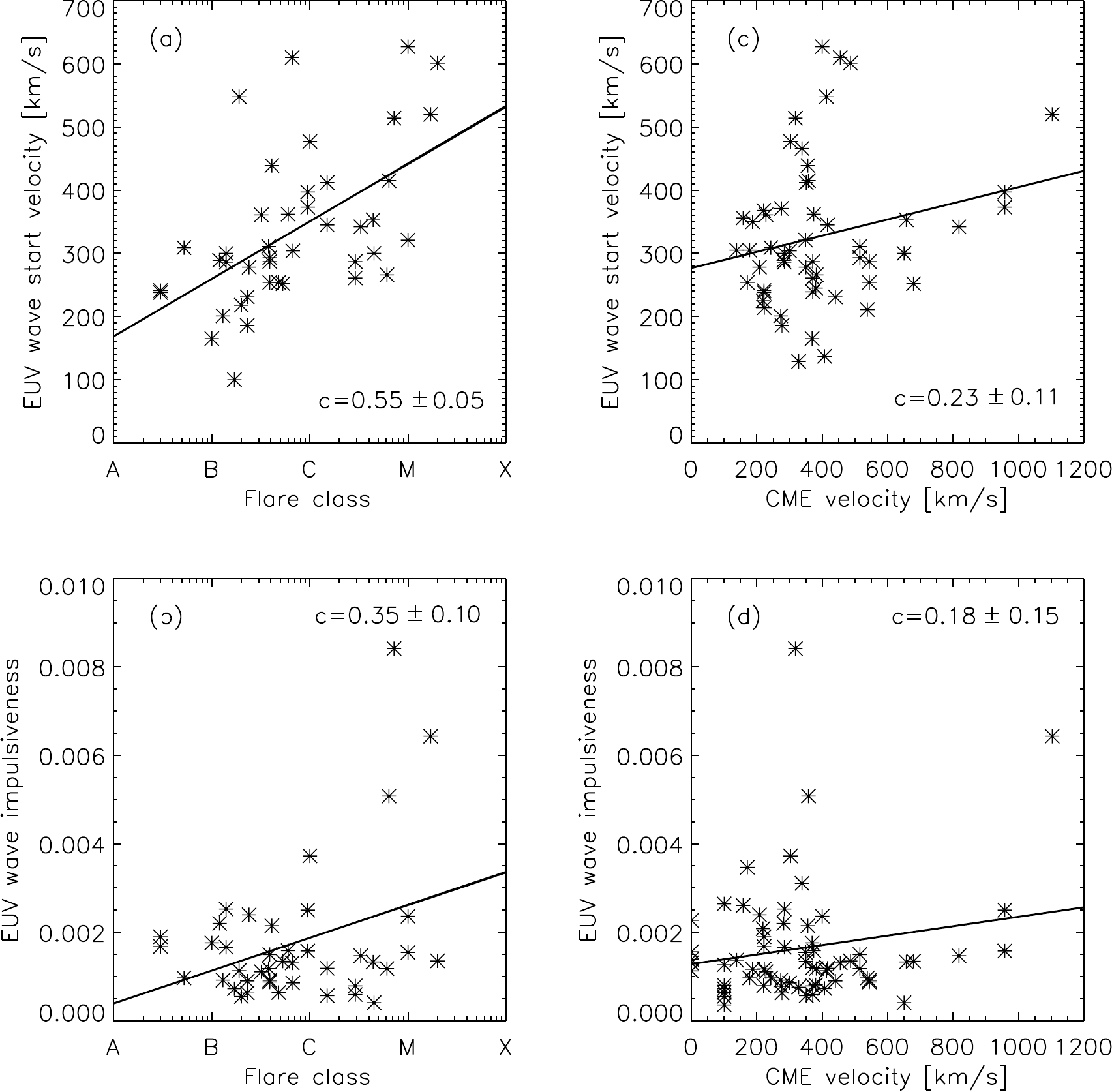}
    \caption[.]{Scatter plots of EUV wave parameters \textit{vs.} flare and CME parameters. Top: wave start velocity \textit{vs.} flare class (a) and CME velocity (c); Bottom: wave impulsiveness \textit{vs.} flare class (b) and CME velocity (d).} \label{img:correlation_muhr2011_2mal2}
\end{figure*}

In Figure~\ref{img:correlation_muhr2011_2mal2} we present the scatter plots of wave velocity and impulsiveness against the flare GOES classification (panels a, b) and against the CME velocities (panels c, d). The correlation coefficients to the CME speeds are relatively low, $c=0.23\pm0.11$ (Figure~\ref{img:correlation_muhr2011_2mal2}(c)). In Figure~\ref{img:correlation_muhr2011_2mal2}(d) we correlated the wave impulsiveness with the CME velocities, finding a correlation coefficient of $c=0.18\pm0.15$. The correlation coefficients for the wave parameters against the GOES soft X-ray flare class are considerably higher than those against the CME speeds. We find correlation coefficients of $c=0.55\pm0.05$ for the wave velocity and $c=0.35\pm0.10$ for the impulsiveness against the flare class.

\subsection{Events Observed by Both ST-A and ST-B}
Eight of the 60 events under study are observable and measurable in
both ST-A and ST-B. These are the events of 16 May 2007, 19 May 2007, 22 May 2007, 7 December 2007, 26 April 2008, 16 May 2008, 21 July 2008, and 27 February 2009. For all eight events the kinematical measurements
from two different vantage points show similar results
within the error limits (see Table~\ref{tab:table1} and~\ref{tab:table2}, bold face). We note that for each of these events the analyzed sector was the same in both ST-A and ST-B. Thus kinematical results are comparable, except for the differences to the extent affected by spatial integration along different lines of sight. The separation angle of the two spacecraft lies for the eight examples in the range from 8$\degr$ to 92$\degr$. Two examples observed in ST-A and ST-B
are shown in Figures~\ref{img:20070516A} and \ref{img:20071207B}. The
first data set shows the event of 16 May 2007 with a separation
angle of 8$\degr$ between the two spacecraft. The measurements for
both spacecraft are: mean velocity $v_{\rm{ST-A}}\approx212\pm6$~km~s$^{-1}$ and $v_{\rm{ST-B}}\approx217\pm5$~km~s$^{-1}$,
start velocity $v_{\rm{ST-A}}\approx271\pm32$~km~s$^{-1}$ and $v_{\rm{ST-B}}\approx318\pm22$~km~s$^{-1}$, and the deceleration
$a_{\rm{ST-A}}\approx-30\pm18$~m~s$^{-2}$ and $a_{\rm{ST-B}}\approx-54\pm13$~m~s$^{-2}$. The second data set shows the event
of 7 December 2007 with a separation angle of $42\degr$. The kinematical
measurements give: mean velocity
$v_{\rm{ST-A}}\approx263\pm10$~km~s$^{-1}$ and $v_{\rm{ST-B}}\approx294\pm11$~km~s$^{-1}$,
start velocity $v_{\rm{ST-A}}\approx299\pm61$~km~s$^{-1}$ and $v_{\rm{ST-B}}\approx326\pm54$~km~s$^{-1}$, and the deceleration
$a_{\rm{ST-A}}\approx-22\pm39$~m~s$^{-2}$ and $a_{\rm{ST-B}}\approx-23\pm40$~m~s$^{-2}$.

The kinematical measurements revealed consistent results (within the error limits) for the observations from the different STEREO vantage points. However, for the amplitude of the disturbance significant differences can result for the ST-A and ST-B observations. In two out of eight events studied in both STEREO s/c, we obtained significant differences in the perturbation amplitude  (2008 April 26; 2008 July 21). Such differences in the morphology are reproduced in MHD simulations of the line-of-sight integrated emission of EUV waves observed from different vantage points \inlinecite{hoilijoki13}.

\section{Discussion and Conclusions}\label{sec:discussion}
We analyzed 60 well pronounced EUV wave events observed by the
STEREO-EUVI telescopes regarding their kinematics, correlations between characteristic parameters of the propagating disturbance, and their association to CMEs, flares, and type II bursts. We stress here that our analysis and the following conclusions refer only to strong EUV waves, since only those had been selected from the total sample.

From the kinematical analysis we obtain for the mean velocity and start velocity arithmetic means of $v\approx254\pm76$~km~s$^{-1}$ and $v\approx312\pm115$~km~s$^{-1}$, respectively. The acceleration values are between $-460$ and $+50$~m~s$^{-2}$ with a mean of $-97\pm103$~m~s$^{-2}$ indicating deceleration during the propagation. Our results for the wave velocities are in accordance with former studies of EUV waves based on observations from the EIT and EUVI instruments \citep{klassen00,biesecker02,thompson09,warmuth11}. We find a strong anticorrelation between the start velocity and the acceleration of the EUV waves,  $c=-0.77\pm0.07$, which indicates that faster wave events decelerate stronger than slower events. A similar result was obtained by \inlinecite{warmuth11} who studied a combined set of EUV waves observed with EIT and EUVI. Our study shows that the linear fit to the scatter plot start velocity against acceleration $a$  of the EUV wave intersects the $a = 0$~m~s$^{-2}$ line (indicating propagation at constant speed) at a velocity of 230~km~s$^{-1}$. This value is well consistent with the fast magnetosonic speed in the quiet solar corona, providing strong support that the EUV waves under study are indeed fast-mode MHD waves. We note that in \inlinecite{warmuth11} three different classes of EUV waves were found. Comparing these two studies reveals that our sample corresponds to their class~1 and~2 events. The events that would belong to class~3 according the \inlinecite{warmuth11} study, are not covered in our sample as we selected only strong and well pronounced EUV waves where we could follow the evolution. Our interpretation that the events under study are fast-mode MHD waves is in accordance with the \inlinecite{warmuth11} interpretation of their class~1 and~2 events.

\begin{figure*}
    \centering%
    \includegraphics[width=0.90\textwidth,keepaspectratio=true]{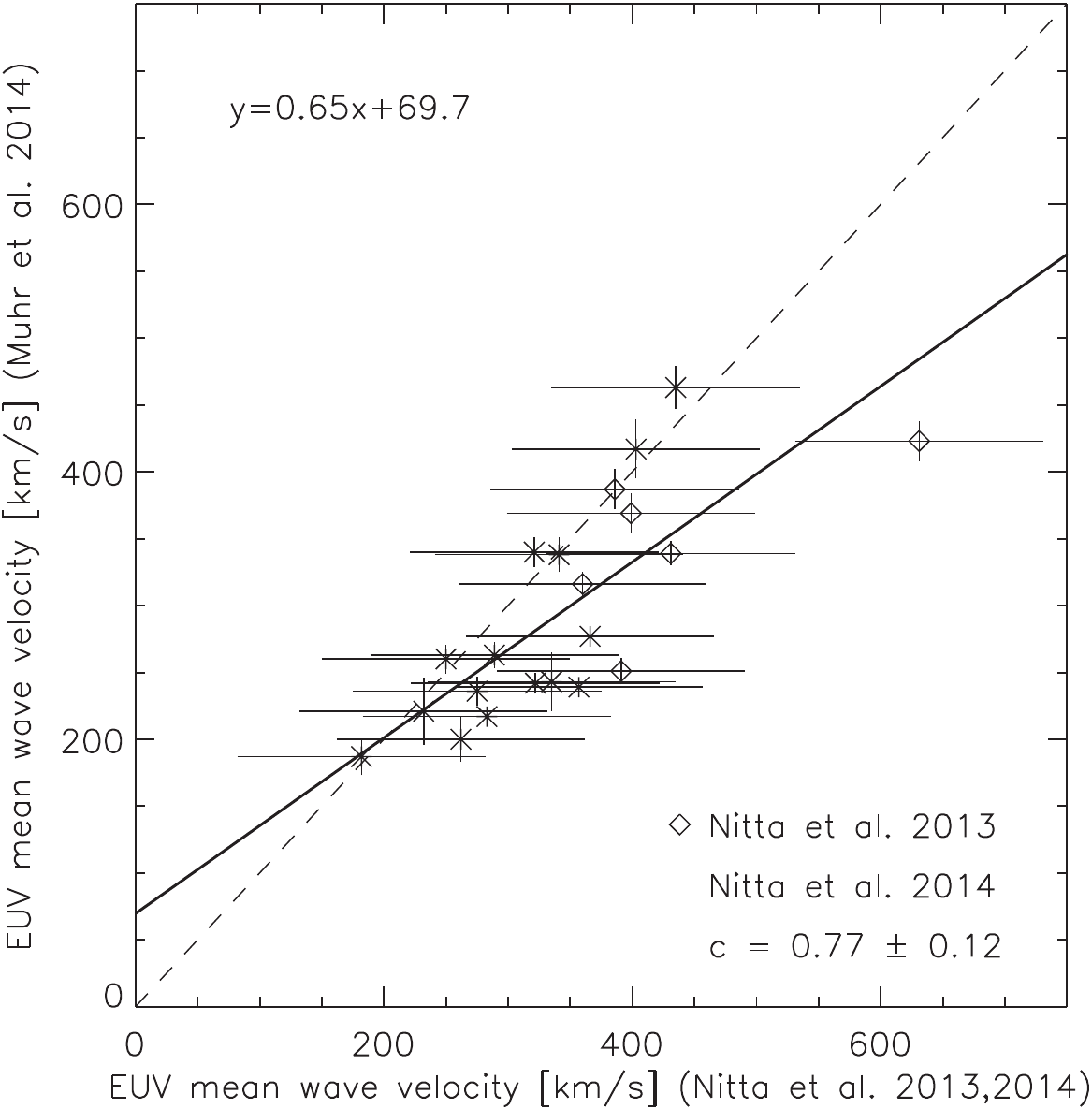}
    \caption[.]{Comparison of our velocity measurements ($y$-axis) against the velocity measurements from \inlinecite{nitta13b} and \inlinecite{nitta14} ($x$-axis). The solid line is a linear regression line to the data points. The fit and the
    correlation coefficients are given in the inset. The dashed line is the one-to-one correspondence.} \label{img:nitta1314}
\end{figure*}

\begin{figure*}
    \centering%
    \includegraphics[width=0.80\textwidth,keepaspectratio=true]{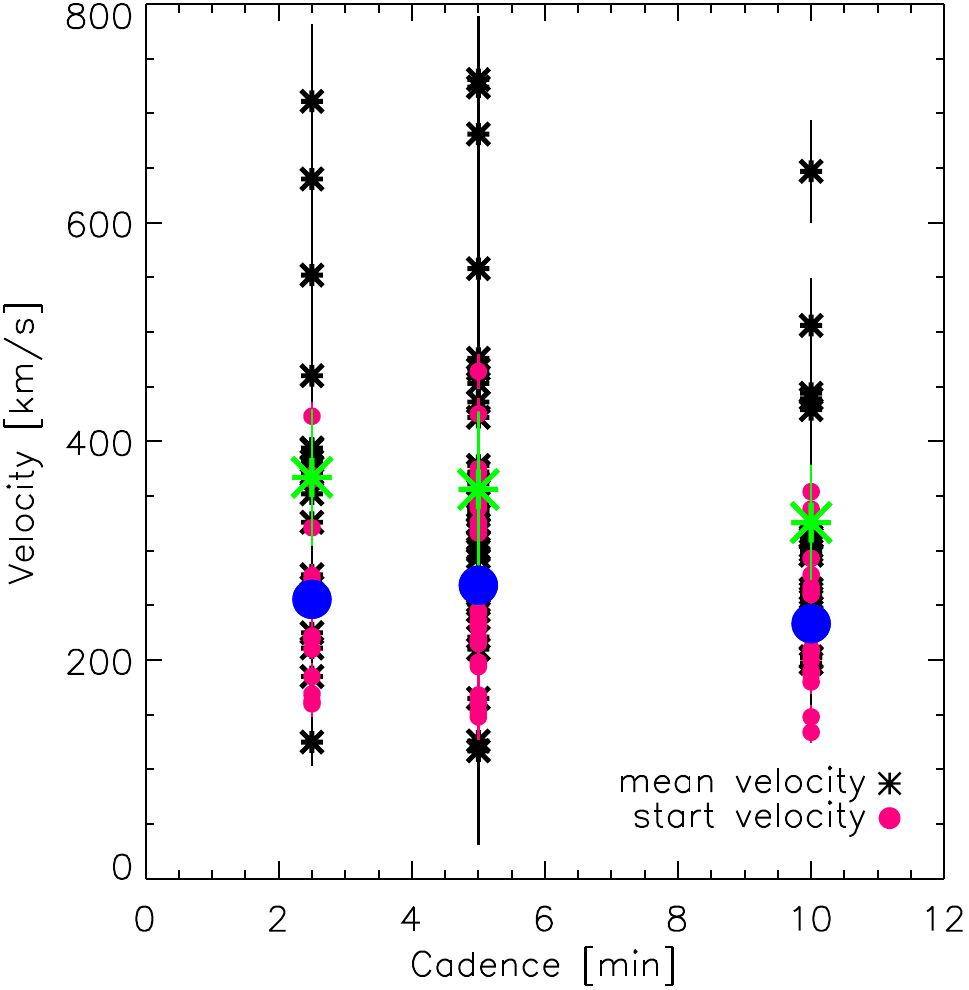}
    \caption[.]{Mean velocities (pink filled circle) and start velocities (black asterisks) of all 60 EUV wave events under study versus the observing cadence (either 2.5~min, 5~min, or 10~min) of each event. The arithmetic mean values for both, the mean velocities (blue filled circle) and the start velocities (green asterisks) are overplotted for each cadence sub-group.} \label{img:vel_cad}
\end{figure*}

A recent study by \inlinecite{nitta13b} based on high-cadence SDO/AIA imagery resulted in a much higher mean velocity of $v\approx 640$~km~s$^{-1}$, and revealed no significant correlation between speed and acceleration of the EUV waves under study. A comparison of six events (marked with a ${\dag}$ in Table~\ref{tab:table4}) which are analyzed by \inlinecite{nitta13b} as well as in our study led to the following findings: For five out of six events the mean velocity was $\approx$30\% higher than the mean velocity from our measurements. In one case the velocities were matching each other well.
In another recent study by \inlinecite{nitta14} the association of solar flares with CMEs as well as their association to EUV wave events was analyzed. Using the same STEREO imagery (from 2007 until 2009) they found 34 EUV events associated with solar flares. We took all 15 events (marked with a ${\dag\dag}$ in Tables~\ref{tab:table1} and \ref{tab:table2}) which are analyzed by \inlinecite{nitta14} as well as in our study for a comparison of the results. In Figure~\ref{img:nitta1314} we display the velocity values obtained by \inlinecite{nitta13b} and \inlinecite{nitta14} against ours together with a one-to-one correspondence line (black dashed line). The correlation coefficient is $c=0.77\pm0.12$ and the mean velocity difference is $\Delta v=56\pm53$~km~s$^{-1}$. What can be seen is that the mean velocity values found by \inlinecite{nitta13b} and \inlinecite{nitta14} are in 50\% of all cases higher than ours. Thus, the mean velocity for these 21 events found by \inlinecite{nitta13b} and \inlinecite{nitta14} is $v=345$~km~s$^{-1}$ while our mean velocity value is $v=296$~km~s$^{-1}$.
An explanation for the higher mean velocities obtained in the \inlinecite{nitta13b,nitta14} studies may be that the velocities were derived during the early propagation phase (until about 15--20 min after launch). This rather short observation time leads to higher velocities since the deceleration occurs only in the later stages of the evolution. Additionally, in some cases the propagation sector used for the analysis is different. Thus the velocities may differ due to the non-isotropic behavior of EUV waves.
\inlinecite{nitta13b} extracted the velocity at the propagation sector where it was highest. \inlinecite{long08} reported for the event of 19 May 2007 that a higher observing cadence in different channels can lead to higher wave velocities. For checking the dependence of the derived results on the time cadence, we plot in Figure~\ref{img:vel_cad} the mean velocities (pink filled circle) and start velocities (black asterisks) of all 60 EUV wave events under study versus the observing cadence (either 2.5~min, 5~min, or 10~min) of each event. The sample average values for both, the mean velocities (blue filled circle) and the start velocities (green asterisks) are overplotted for each sub-group. The arithmetic means for the mean velocities are: $v=256\pm11$~km~s$^{-1}$ (for 2.5~min), $v=268\pm14$~km~s$^{-1}$ (for 5~min), and $v=233\pm12$~km~s$^{-1}$ (for 10~min) and for the start velocities are: $v=367\pm63$~km~s$^{-1}$ (for 2.5~min), $v=356\pm72$~km~s$^{-1}$ (for 5~min), and $v=326\pm53$~km~s$^{-1}$ (for 10~min). Thus, there is no trend observable that the difference in the observing cadence (from 2.5 to 12 min) causes systematic differences in the EUV wave velocities derived. However, we cannot exclude that there is an effect when the cadence is as small as 12~s (AIA) and the wave velocity is determined in the very early propagation phase of the event, such as done in the \inlinecite{nitta13b} study.

We find positive correlations between the velocity and the peak amplitude as well as the impulsiveness of the perturbation which supports the interpretation that the EUV waves are pressure pulses that compress the surrounding plasma while propagating. The anticorrelation between the velocity and the build-up time means that for impulsive events the build-up time is shorter than for gradual events.
Thus, the steepening of the peak amplitudes into a shock front depends on the initial velocity and acceleration, which is a typical characteristic of wave-like disturbances \citep{vrsnak00}.

The association rates of the events under study to CMEs and flares may give hints to the driver of the transients. For the 60 events under study we find 57 associated CMEs giving a 95\% association rate. This association rate is similar to numbers published by \inlinecite{biesecker02}. However, there is only a very weak correlation between the CME velocities and the characteristic parameters of the EUV wave ($c\approx0.2$). A loose correlation between CME and EUV speeds is also reported by \inlinecite{nitta13b} from a SDO/AIA statistical study. They found that EUV wave events with no CME correlation show smaller velocities compared to EUV wave events associated with CMEs. An explanation for this low correlation may be that in the early evolution of CMEs the expansion in lateral as well as radial direction (often noticed as spherical CME bubble) is basically equally strong, whereas later on the radial evolution is dominant and the lateral weakens/finishes (limited width of a CME versus radial evolution into IP space). Thus, further out we measure more or less the radial evolution, hence, the coronagraphic measurements are biased and show a loose relation to EUV waves. Additionally, it is the main acceleration phase of the CME often occurs below $0.5$\textit{R}$\sun$ \citep{temmer08,temmer10,bein11} that is probably the relevant factor for the EUV wave generation. In this comparison, one has to consider that the extracted LASCO/CME mean speeds are projected speeds they reflect the propagation behavior at distances far from the Sun ($r>2$\textit{R}$\sun$). Detailed case studies of limb events including multipoint STEREO observations have shown that there is a close relation between the CME expanding flanks and the EUV wave, indicating that the fast expanding CME flanks initiate the EUV wave \citep{patsourakos09,kienreich09,veronig10,liu12,cheng12}.

The flare association rate is considerably lower with a value of $\approx$74\%, similar to the findings of \inlinecite{klassen00}, \inlinecite{biesecker02}, and \inlinecite{cliver05}. As we are analyzing the events occurring during the rising phase of solar cycle 24  most of the associated flares are of SXR class A to C, and may thus not be energetic enough to act as driving agent. Nevertheless, those events associated with a solar flare lead to a relatively good correlation coefficient of $c=0.55\pm0.05$ between the GOES flare class and the EUV wave velocities suggesting that both phenomena are related to each other. A similar trend of a loose correlation between the flares' SXR class and EUV speeds has also been noted by \inlinecite{nitta13b}. They found that those wave events with higher speeds are accompanied by more intense flares. This can be explained, as the more energy released in an active region the more powerful a flare/CME event can become and thus the stronger the push to the surrounding corona which can generate a stronger and faster EUV wave event.

We note that those events without a CME association also lacked a solar flare association, namely the events of 16 May 2008, 21 July 2008, and 26 April 2009. In these cases neither the CME nor the flare seem to be responsible for the initiation process. \inlinecite{zheng12} studied an EUV wave without CME, concluding that it was driven by the associated jet. These observations support the scenario discussed by \inlinecite{vrsnak08}, in which the deciding factor for the launch of an EUV wave or a coronal shock wave is the impulsive mass motion perpendicular to the magnetic field. This mass motion may be related to a CME escaping to interplanetary space, but it might be also confined over a certain distance range in the solar atmosphere. In addition to the CME and flare association rate we investigated the type II burst association rate and found a rate of 22\% for coronal and 15\% for interplanetary type II bursts, indicating shock formation in the solar corona and in interplanetary space, respectively. This is a similar outcome to the study by \inlinecite{biesecker02} with an association rate of EUV waves to coronal type II bursts of 23\%.

Events observed in both ST-A and ST-B are of special interest as there is criticism \citep{hoilijoki13} whether kinematical results obtained from different vantage points are even comparable. In our study 8 out of 60 events were simultaneously observed from two vantage points. Whereas in the beginning of the STEREO mission the separation angle between the two s/c was small with $\approx$8$\degr$, the separation angle of $\approx$45$\degr$ in the early months of 2008 clarified the situation. For seven out of eight events the kinematical results from both spacecraft were consistent with each other within the uncertainties, similar to the case study by \inlinecite{temmer11}. For the event of 21 July 2008 the mean velocities obtained differ by about 20\% for ST-A and ST-B, whereas the start velocities are similar,  $v_{\textrm{ST-A}}\approx$375$\pm$75~km~s$^{-1}$ and $v_{\textrm{ST-B}}\approx$357$\pm$109~km~s$^{-1}$. This difference in the mean velocity is due to the fact that in the ST-A observations more data points could be used, resulting in a longer observation time than for ST-B. Since the deceleration happens during the later stage the mean linear velocity is smaller for ST-A than for ST-B. These results indicate that the EUV wave kinematics is not dramatically affected by the different vantage points. However, for two events (16 May 2008 and 21 July 2008), where the s/c separation is comparatively large (70$\degr$ and 90$\degr$), we found differences in the morphology and the perturbation amplitudes (of $\approx$30\%) for ST-A and ST-B. This implies that the CME plasma can also contribute to the line-of-sight integrated optically thin emission across the wave front which highly depends on the geometry with respect to the observer as suggested by \inlinecite{ma09}, \inlinecite{patsourakos09} and \inlinecite{hoilijoki13}. These findings suggest that the perturbation amplitude may indeed be considerably affected by line-of-sight effects, and has thus to be interpreted with caution.

In conclusion, we find for the events under study a broad velocity range from close to the local fast magnetosonic speed $v_{\rm{fm}}$ up to remarkably higher velocities, a strong anticorrelation between acceleration and start velocities with the results that initially faster events decelerate stronger and events with zero acceleration propagate at nearly $v_{\rm{fm}}$. This pronounced anticorrelation, the steepening, broadening and decaying of the transients' perturbation profiles as well as correlations between the propagation velocity and perturbation profile characteristics (\textit{e.g.} amplitude, impulsiveness) provide strong evidence that EUV waves are indeed fast-mode MHD waves.

%
 \begin{acks}
N.M., I.W.K., A.M.V. and M.T.
acknowledge the Austrian Science Fund (FWF): P20867-N16, P-24092-N16 and V195-N16. N.M. acknowledges the
MOEL-Plus F\"orderprogramm. This work has received funding from the European Commission FP7 Project no. 284461 [eHEROES].
The STEREO/SECCHI data are produced by an international consortium
of the Naval Research Laboratory (USA), Lockheed Martin Solar and
Astrophysics Lab (USA), NASA Goddard Space Flight Center (USA),
Rutherford Appleton Laboratory (UK), University of Birmingham
(UK), Max-Planck-Institut f\"ur Sonnensystemforschung (Germany),
Centre Spatiale de Li$\grave{e}$ge (Belgium),
 Institut d'Optique
Th$\acute{e}$orique et Appliqu$\acute{e}$e (France), and Institut
d'Astrophysique Spatiale (France).
 \end{acks}


\appendix
\section{Appendix}
\subsection{Comparison of the Visual Tracking and the Perturbation
Profile Methods}\label{sec:results_correlation} 
\begin{figure*}[h]
    \centering%
    \includegraphics[width=1.00\textwidth,keepaspectratio=true]{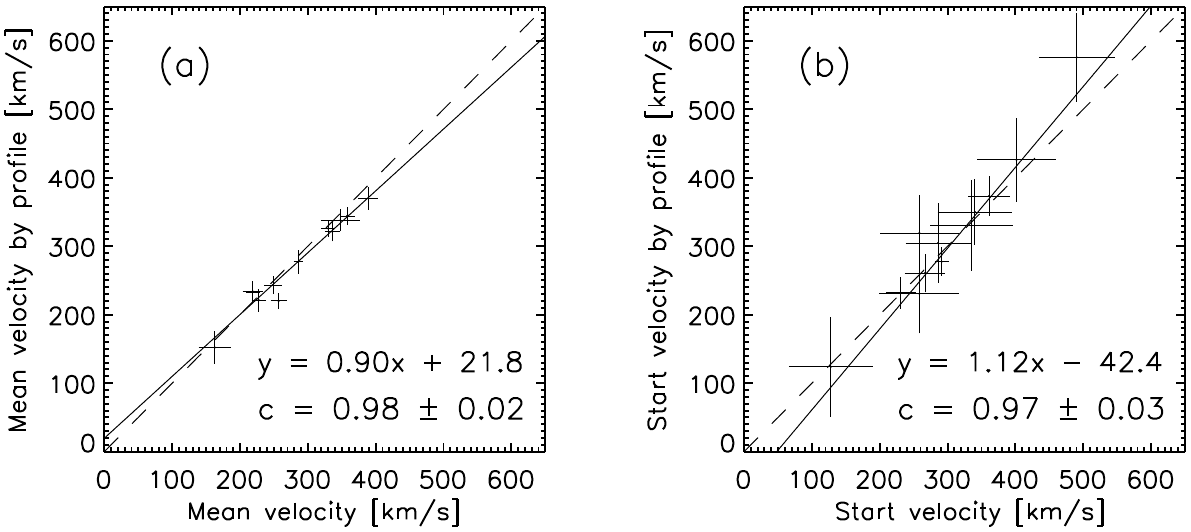}
    \caption[.]{Comparison of visual tracking and perturbation profile method.
    In the first panel the mean velocities of a representative subsample of 12 EUV waves derived from the visual tracking method ($x$-axis)
    are plotted against the mean velocity values derived from the perturbation profiles ($y$-axis).
    In the second panel the starting velocity values derived from both methods are plotted against
    each other. The solid lines are linear regression lines to the data points. The fit and the
    correlation coefficients are given in the inset. The dashed line is the one-to-one correspondence.} \label{img:correlation}
\end{figure*}

\inlinecite{muhr11} and
\inlinecite{long11} present a comparison of the two basic methods that are used to derive EUV wave kinematics, namely visual tracking of the wave fronts and perturbation profiles, finding good consistency in the obtained results. However, both studies were using a small sample of four EUV
wave events. Here we study a representative sub-sample of 12
events on which we apply the visual as well as the perturbation
profile method to a series of BR images in order to compare both kinematical analysis techniques. For both methods we focused on the same specific
45$\degr$ sector in which the disturbances show highest contrast
levels. In Figure~\ref{img:correlation} we display the correlation of the
derived velocity values of the 12 EUV wave events. In the two panels
the x-axes represent the velocities received by the visual tracking
method while the y-axes represent the velocities received by the
perturbation profile method. Figure~\ref{img:correlation}(a) shows
the mean velocities and a linear fit to the data with a linear regression line of $y=0.90x+21.8$ and a correlation
coefficient of 0.98$\pm$0.02. Figure~\ref{img:correlation}(b) shows the
start velocities at the first wave occurrences derived from the quadratic fits to the time-distance curves and a linear fit to the data with a linear regression line of $y=1.12x-42.4$ , giving a correlation
coefficient of 0.97$\pm$0.03. The high correlation coefficients and the linear fit parameters derived show that both methods give consistent results for the velocities describing the EUV wave propagation.

\newpage

\newpage

\newpage

\newpage

\newpage

\newpage

\newpage

\newpage

\end{article}
\end{document}